\documentclass[12pt,preprint]{aastex}
\usepackage{mathrsfs}
\usepackage{float}
\usepackage{bm}
\usepackage{color}
\usepackage{ulem}

\newcommand{\bc}{}
\newcommand{\rc}{}

\shorttitle{Terrestrial-Mass Planets in the HZ of M Dwarfs.}
\shortauthors{Owen, J. E., Mohanty, S.}

\begin{document}

\def\ts{T_{\rm surf}}
\def\rhos{\rho_{\rm surf}}
\def\ps{P_{\rm surf}}
\def\tbb1{T_{\rm BB,MS}}
\def\msun{M$_{\odot}$}
\def\lsun{L$_{\odot}$}
\def\me{M$_{\oplus}$}
\def\re{R$_{\oplus}$}
\def\mfraci{m_{\rm frac,i}}
\def\tcooli{t_{\rm cool,i}}
\def\mcore{M_{core}}
\def\lbol{$L_{\rm bol}$}
\def\teff{$T_{\rm eff}$}

\title{Habitability of Terrestrial-Mass Planets in the HZ of M Dwarfs.\\ I. H/He-Dominated Atmospheres. }

\author{James E. Owen\altaffilmark{1,2}, Subhanjoy Mohanty\altaffilmark{3}}

\altaffiltext{1}{Institute for Advanced Study, Einstein Drive, Princeton NJ, 08540, USA. {\it jowen@ias.edu}}
\altaffiltext{2}{Hubble Fellow}
\altaffiltext{3}{Department of Physics, Astrophysics Group, Imperial College London, 1010 Blackett Laboratory, Prince Consort Road, London SW7 2AZ, UK. {\it s.mohanty@imperial.ac.uk}}

\begin{abstract}

The ubiquity of M dwarfs, combined with the relative ease of detecting terrestrial-mass planets around them, has made them prime targets for finding and characterising planets in the ``Habitable Zone'' (HZ). However, {\it Kepler} finds that terrestrial-mass exoplanets are often born with voluminous H/He envelopes, comprising mass-fractions ($M_{env}/M_{core}$) $\gtrsim 1$\%. If these planets retain such envelopes over Gyr timescales, they will not be ``habitable'' even within the HZ. Given the strong X-ray/UV fluxes of M dwarfs, we study whether sufficient envelope mass can be photoevaporated away for these planets to become habitable. We improve upon previous work by using hydrodynamic models that account for radiative cooling as well as the transition from hydrodynamic to ballistic escape. Adopting a template active M dwarf XUV spectrum, including stellar evolution, and considering both evaporation and thermal evolution, we show that: {\it (1)} the mass-loss is {\rc (considerably)} lower than previous estimates that use an ``energy-limited'' formalism and ignore the transition to Jeans escape; {\it (2)} at the inner edge of the HZ, planets with core mass $\lesssim 0.9$\,\me\, can lose enough H/He to become habitable if their initial envelope mass-fraction is $\sim$1\%; {\it (3)} at the outer edge of the HZ, evaporation cannot remove a $\sim$1\% H/He envelope even from cores down to 0.8\,\me. Thus, if planets form with bulky H/He envelopes, only those with low-mass cores may eventually be habitable. Cores $\gtrsim$ 1\,\me, with $\gtrsim$1\% natal H/He envelopes, will {\it not} be habitable in the HZ of M dwarfs.

\end{abstract}


\section{Introduction}

M dwarfs comprise the bulk ($\sim$75\%) of the stellar population of our galaxy.  Moreover, their low masses and small radii, compared to Sun-like stars, make planets of a given mass, size and orbital separation much easier to detect around them via both the transit and Doppler methods. In addition, the Habitable Zone (HZ) of such low luminosity red dwarfs lies considerably closer to the central star than in the case of solar-types; this further enhances the transit and Doppler signatures of HZ planets, and also allows a larger number of planetary orbits to be observed in a given time, making the detection and characterisation of such planets even easier.  Consequently, the next generation of missions investigating exoplanets are aimed at later-type stars rather than the mainly solar-type ones targeted by {\it Kepler}, and discussions of exoplanet habitability increasingly focus on M dwarf systems. 

There are many complications pertaining to the habitability of planets around M-dwarfs, even if they possess surface temperature and pressure conditions favourable to liquid water, such as tidal locking \citep[e.g.][]{heller11,leconte15}, run-away greenhouse effects \citep[e.g.][]{kopparapu2013} and water loss \citep{luger15}. The first major uncertainty, however, is whether they can possess such conditions at all, given what we currently know about exoplanets {\bc \citep[e.g.][]{scalo07}}.    

The last decade of exoplanet discoveries has dramatically altered our understanding of a ``standard'' low-mass planet. {\it Kepler} has detected thousands of planetary candidates \citep[e.g.][]{mullally2015}, the majority of which are small \citep[$\lesssim 3$~R$_\oplus$:][]{howard12,petigura2013,morton14,silburt2015} and close to their host stars. Correcting for observational biases, one finds that most stars probably harbour at least one such ``Kepler'' planet \citep{fressin13}. Masses have been now been obtained for a significant number (albeit a small fraction) of these exoplanets, by combining {\it Kepler} transit data with either transit timing variations \citep[TTVs; e.g.,][]{wu13,hadden14,jontofhutter14,jontofhutter15} or radial velocity measurements \citep[e.g.,][]{weiss14,marcy14,dressing15}. A very surprising result to emerge from this work is that a large fraction of low-mass, close-in Kepler planets are enshrouded by voluminous H/He envelopes, which contain a non-negligible fraction of the planet's mass \citep[as demonstrated, for instance, by the extremely low densities of some: $\lesssim 1$~g~cm$^{-3}$; e.g.,][]{wu13,jontofhutter14,jontofhutter15}: very unlike the low-mass planets in our own solar system. 

In particular, statistical analyses of the properties of the low-mass exoplanet population reveal that the latter are inconsistent with a completely rocky composition \citep{rogers15,wolfgang15b}. Instead, comparisons of the {\bc observed exoplanets} with measured masses and radii to structural models \citep{wolfgang15a} indicate that the dominant structure is a solid core overlaid with a $\sim$1\% (by mass) H/He envelope. Since close-in planets are subject to intense irradiation by high energy stellar photons, which can drive an outflow by heating the upper layers of the atmosphere to close to the escape temperature \citep[e.g.][]{lammer03,tian05,mc09,oj12,erkaev16}, the evaporation of such H/He envelopes becomes important for the planets' evolution \citep[e.g.][]{baraffe05,lopez12}. Forward modelling studies suggest that the majority of close-in exoplanets were born with H/He envelopes, but about half have subsequently lost these through evaporation \citep{ow13,lopez13}. In summary, current exoplanet demographics lead us to infer that a dominant mode of low-mass planet formation produces a rocky Earth-like core surrounded by a H/He envelope with a mass-fraction of $\sim$1\% \citep{wolfgang15a}. This observational result is supported by recent theoretical calculations within the framework of core accretion, which suggest that low-mass cores $<5\,$M$_\oplus$ will acquire envelope mass-fractions of order a few percent \citep{rogers11,bodenheimer14,lee14,lee15}. For example, the scalings of \citet{lee15} imply that a 1\,M$_\oplus$ core will accrete a H/He envelope mass-fraction of 0.1-1\% during the gas disk's lifetime. 

{\rc We note that various studies of gas accretion onto Earth-like cores find that the results are sensitive to a number of uncertain parameters, such as the opacity and disc properties \citep[e.g.][]{ikoma12,lammer14,stoekl15}, and, especially, the assumed core accretion rate. Thus, these studies suggest a range of initial H/He envelope masses for a given core mass. However, this theoretical prejudice should not blind us to the {\it empirical} fact that current exoplanet data indicate that envelope mass fractions $\sim 1$\% are preferred\footnote{see \citet{ow15} for a discussion of why much larger envelopes may be scarce, as the data suggest.}. }

For a terrestrial-mass planet, orbiting either a solar-type star or an M dwarf and located within the classical HZ \citep[defined here as the range of orbital separations where an Earth-mass planet, with roughly Earth-like composition and atmosphere, can harbour liquid surface water; eg., see][]{kopparapuetal2013}, a H/He envelope with mass-fraction of order a percent would preclude habitability, by yielding very high surface temperatures and pressures incompatible with liquid water. However, habitable conditions may be achieved if evaporation can reduce the H/He envelope mass-fraction to $\ll 10^{-3}$ \citep[e.g.][]{pierrehumbert11}; strip away this primordial envelope entirely so that it is replaced by a tenuous secondary atmosphere, such as those of the solar system terrestrial planets \citep[e.g.][]{kasting93,kopparapuetal2013}; {\bc or separate H and He in planets with a low initial H/He fraction to leave a habitable He atmosphere \citep{hu15}}. In other words, evaporation may turn a large population of uninhabitable low-mass planets with voluminous H/He envelopes into habitable ones. Such an effect is unlikely in the HZ of Sun-like stars, since their X-ray/UV fluxes over Gyr timescales at $\sim$1\,AU are too low to remove a {\bc massive} H/He envelope {\rc \citep[though several works have shown it is possible to remove a {\it small} H/He atmosphere; e.g.,][]{tian2005b,erkaev13,lammer14}}. M dwarfs, on the other hand, are far more active, and remain so for much longer, than solar-types, leading to much higher XUV fluxes within their HZs over Gyr timescales \citep[e.g.][]{guedel04,jackson12}. Moreover, low-mass planets appear abundant around these red dwarfs, which are themselves the most common stars in the galaxy; evaporation, if efficient enough, could thus lead to a plethora of habitable planets in M dwarf systems. 

In this work, therefore, we consider terrestrial-mass planets in the HZ of M dwarfs, with an initial H/He envelope mass-fraction of $\sim$1\%, and investigate whether evaporation over a Gyr can remove a sufficient portion of this atmosphere to render the planet habitable at the end. We account for both stellar evolution and the thermal evolution of the planet. In order to make advances in this important area, we also abjure various simplifying -- and, as it turns out, erroneous -- assumptions about the evaporative flow made in previous studies of this subject, as described below, in order to obtain a more rigorous and realistic estimate of the mass loss.

\section{Overview}

\subsection{Comparison to Previous Work}

Given the {\it Kepler} results above, and the interest in habitable planets, several recent studies have investigated the evaporation of voluminous H/He envelopes around terrestrial-mass solid cores. However, previous studies have typically made two key simplifying assumptions: {\it (i)} radiative cooling in the flow is either neglected, or accounted for by assuming a fixed energy efficiency for driving the flow - either in a global or local sense; and {\it (ii)} evaporation is assumed to occur in the hydrodynamic limit at all times. 

Both assumptions are likely to lead to an overestimation of the amount of H/He a planet can lose. First, radiative cooling becomes important when the timescale for a fluid element to advect its heat outwards (the `flow timescale') becomes comparable to the timescale for radiative losses to cool that fluid element \citep[e.g.][]{oa15}. This lowers the temperature in the flow, which in turn pushes the sonic point to greater heights (lower densities) and thus reduces the mass-loss rate. Second, in order to be in the hydrodynamic limit, one requires that the gas remain collisional up to the sonic point \citep[e.g.,][]{oj12}. If this is not satisfied, the flow will collapse\footnote{There may be a small transition region where mass loss occurs sub-sonically \citep{tian05}; however, these ``breeze'' solutions are exponentially sensitive to density and possibly unstable \citep{velli94}.} and switch to Jeans escape, which has a much reduced mass-loss rate (as we demonstrate later, Jeans escape cannot remove a significant H/He atmosphere on Gyr time-scales, and other non-thermal processes are also unlikely to play a significant role). Finally, these two points are inter-related: as radiative cooling pushes the sonic point to higher heights and lower densities, it can also trigger the transition to Jeans escape earlier than in calculations that do not include cooling. 


{\bc \citet{lammer14} and \citet{johnstone15} studied the evolution of terrestrial mass cores in the HZ of solar-type stars, using evaporation rates calculated in the ``locally energy-limited'' approach, where a fixed fraction of the absorbed photons' energy is locally deposited into heat. {\rc This fraction is calibrated to detailed calculations by \citet{shematovich14} that explicitly solve the micro-physics of photon absorption; as such, it is more accurate than the standard ``energy-limited'' formalism, wherein an ad hoc fraction of the total incoming radiative flux goes into heating the gas. However, these calculations still neglect radiative cooling, and as such will overestimate the mass-loss rate when radiative cooling is important\footnote{{\rc \citet{shematovich14} caculate the fraction of the incoming XUV flux that goes into heating the gas (via Coulomb collisions of gas particles with photoelectrons liberated by the XUV), instead of being diverted into exciting, ionising or dissociating the atomic and molecular species. However, their calculation does not address radiative cooling; as such, this fraction does not equal the {\it net} heating rate, which is the difference of the heating and cooling rates, and the quantity of importance here.}}}. \citet{lammer14} {\rc further} argue that if the core accretion rate is very large (mass doubling times $\sim 10^5$~years), {\rc resulting in a high-entropy bloated atmosphere, evaporation may be able to completely remove initially massive H/He envelopes from low-mass cores} and leave behind a potentially habitable planet. However, such large core accretion rates require that the time of planet formation to be fine-tuned to occur just before disc dispersal, otherwise the core would end up with a mass $\gtrsim10$~M$_\oplus$.  With more reasonable accretion rates, they find cores are likely to retain a large fraction of their original envelope. \citet{johnstone15}, expanded on this work to demonstrate that  {\rc Earth-mass planets at 1\,AU around solar-type stars can lose a massive 1\% H/He envelope over a Gyr only if the star is an unusually fast rotator (i.e., unusually active, with X-ray luminosity in the upper 90th percentile of the observed spread in X-ray luminosities in these stars). In the majority of cases, i.e., for solar-type stars with more standard rotation rates and actitivity levels, they find  that such planets can only lose H/He envelopes with initial mass-fractions $\lesssim$\,0.1\%}. }  


{\rc \citet{luger15a} have recently investigated photoevaporation around M dwarfs.} {\bc Using the {\rc standard} ``energy-limited'' formalism and further assuming a hydrodynamic flow at all times, they argue that  evaporation can completely strip H/He envelopes with a mass-fraction $\gtrsim 1$\% from Earth-mass cores in the HZs of M dwarfs. If true, this would imply a potentially vast number of worlds with habitable conditions around M dwarfs. However, the problems noted above with the adopted assumptions call this result into question. Specifically, below an envelope mass-fraction of $\sim$1\%, the radius of the planet is an extremely strong function of the remaining envelope mass-fraction \citep{lopez14}; at the same time, the flow timescale increases strongly with decreasing planetary radius (as the atmosphere falls deeper into the planet's gravity well). Thus, one expects radiative cooling to increase significantly as the H/He mass-fraction descends below $\sim$1\%, resulting first in suppressed hydrodynamic mass-loss rates, and then a transition to Jeans escape with greatly diminished loss rates. This process can quench mass loss from planets with a hefty H/He envelope still remaining; indeed, the effect has already been identified at closer separations and higher planetary masses \citep{ow13,lopez13}, and is the origin of the so-called ``evaporation valley'' \citep{jin14}.}

In this paper, therefore, we eschew the assumptions made in earlier studies. Instead, we explicitly account for the effect described above, by: {\it (i)} smoothly transitioning from the regime where heating is balanced by outflow \citep[as in the ``locally energy-limited'' calculations of][]{lammer13} to the regime where heating is balanced by radiative cooling, using a parametrisation calibrated to detailed Monte-Carlo radiative transfer simulations; and {\it (ii)} using our hydrodynamic models to determine when the transition to Jeans escape occurs, triggering a strong suppression in mass-loss rates. In general, with our more appropriate treatment, we find that whether or not a planet can lose enough H/He to be considered habitable depends strongly on the core mass and orbital separation. In particular, the mass loss we derive for $\sim$1\,\me\, cores with initially $\sim$1\% He/He envelopes in the HZ of M dwarfs is much lower than previous estimates, implying that such planets will not be habitable.

\subsection{Model Outline}\label{sec:basics}

Stellar XUV-driven evaporation is particularly important around M dwarfs because these stars are extremely active: for instance, an average early- to mid-M dwarf remains at saturated levels of coronal activity, with $L_X$/\lbol\, $\approx$ 10$^{-3}$--10$^{-4}$, for a few 100\,Myr, compared to only a few$\times$10\,Myr of saturated activity in an {\rc average} solar-type star \citep[e.g.,][]{guedel04}\footnote{{\rc These timescales can be extended, for both M dwarfs and solar-type stars, if the star is initially a very rapid rotator; e.g. \citet{johnstone15,tu15}. The point is that in general M dwarfs evince saturated activity for far longer than solar-types.}}. Consequently, a planet with a given equilibrium temperature receives orders of magnitude more high-energy radiation, integrated over its lifetime, around an M dwarf than around a star like the sun. 

In this study, we are concerned with the evaporation of low-mass planets. \citet{oj12} demonstrated that at high X-ray fluxes, mass-loss from such planets is predominantly X-ray-driven, and EUV heating can be neglected. We thus confine ourselves to X-ray-driven flows here. To study the latter, we need a fiducial X-ray spectrum; we use that of AD Leo, an M3.5 dwarf of mass $\sim$0.4\,\msun, and one of the most active nearby stars.  

Th very strong coronal activity on AD Leo arises not because it is anomalous, but simply because it is relatively young compared to most nearby field stars. At a median age of $\sim$100\,Myr \citep[estimated age range $\sim$25--300\,Myr;][]{shkolnik09}, it is still in its saturated phase of activity, with $L_X$/\lbol\, = 10$^{-3.02}$ \citep[$L_X$ = 7$\times$10$^{28}$\,erg\,s$^{-1}$ and \lbol = 9$\times$10$^{31}$\,erg\,s$^{-1}$;][]{delfosse98}. Its high levels of activity, combined with its proximity ($d \approx 4.9$\,pc), have made AD Leo a touchstone for understanding the coronae of M dwarfs.    

Nevertheless, the intrinsic faintness of M dwarfs (AD Leo included: \lbol\,$\approx 2.3\times$10$^{-2}$\,\lsun) means that, even when $L_X$/\lbol\, is very high, the star is still very faint in X-rays, and it is not yet possible to acquire a full high-quality X-ray spectrum. Instead, one usually reconstructs the spectrum using the technique of Emission Measure Distributions (EMDs), based upon observations of bright emission lines. We use an EMD-derived synthetic X-ray spectrum of AD Leo supplied by J. Sanz-Forcada (pvt.\,comm.\,2014). The methodology for constructing it is described in detail by \citet{sanzforcada11} and \citet{chadney15}; this spectrum is also used by \citet{chadney15}, who show that it agrees very well with current X-ray (and UV) data for AD Leo. We plot the spectrum in Fig.\,1, and compare it to a scaled-solar spectrum (representing a young solar-type star); the comparison solar-like spectrum was calculated in \citet{ercolano08} using the method of \citet{drakecode}, and was designed to be representative of a young, saturated solar-type star. It is immediately clear that the AD Leo spectrum is considerably softer; thus, since X-ray heating is mainly due to photo-electrons liberated by soft $\sim$0.1-1\,keV photons, we expect significantly more heating using our realistic M dwarf spectrum than with a scaled-solar proxy for it \citep[e.g., as done in][]{oj12}.     

\begin{figure}
\centering
\includegraphics[width=\columnwidth]{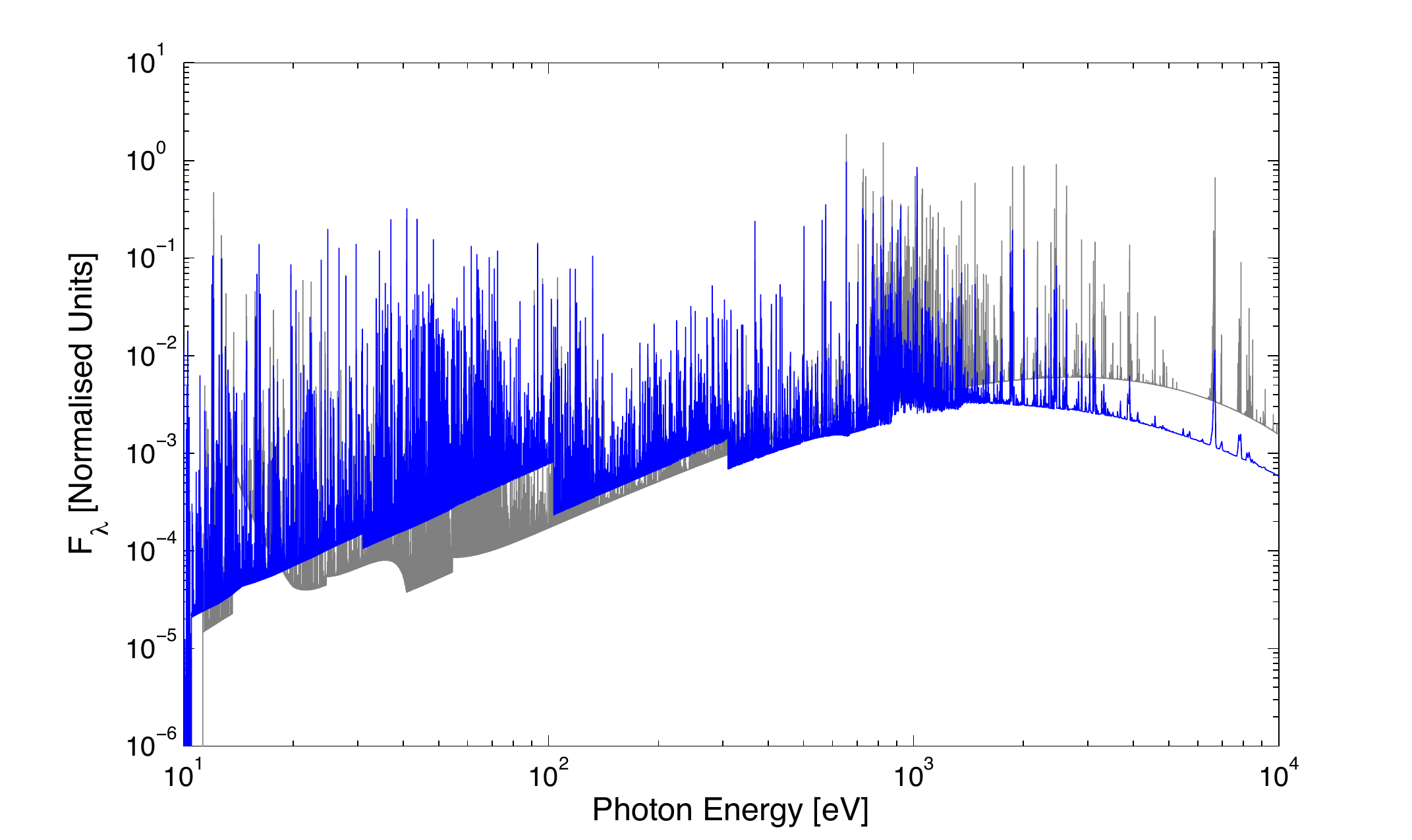}
\caption{Synthetic high-energy spectrum of AD Leo ({\it blue}) compared to that of a young solar-type star calculated by \citet{ercolano08} ({\it grey}). Both spectra are normalized to the same integrated luminosity. Note the higher flux of softer ($\sim$0.1--1\,keV) X-ray photons in AD Leo relative to the young solar-type star.} \label{fig:spec}
\end{figure}

Since AD Leo is currently still in the saturated regime, we assume it has maintained the same $L_X$/\lbol\, = 10$^{-3.02}$ since birth to the present (the standard behaviour of saturated low-mass stars, before stellar rotation slows sufficiently for the activity to become unsaturated). However, its bolometric luminosity certainly {\it has} evolved with time, as the star has contracted towards the Main Sequence; we assume that its \lbol\, follows the theoretical evolutionary track of \citet{baraffe98} for a 0.4\,\msun\, star, and scale $L_X$ accordingly in time. Finally, for ease of computation, we assume that AD Leo will remain saturated, i.e., its $L_X$/\lbol\, will stay unchanged, till 1\,Gyr (when we terminate our evaporation calculations). This is not physically strictly accurate, since a 0.4\,\msun\, star should start spinning down after a few 100\,Myr \citep[e.g.,][]{reiners12}, thus entering the unsaturated regime and evincing decreasing $L_X$/\lbol; assuming saturation at these late times then formally implies that our mass-loss rates will be overestimated here. However, as we shall see, for the core-mass and H/He envelope mass-fraction regime we study in this paper, our calculations show that the planets are either completely stripped of their envelopes well before 100\,Myr, or have entered the Jeans escape regime, with tiny mass-loss rates, by this age. As such, our simplifying assumption of saturated activity beyond a few 100 Myr has no discernible impact on our results.     

In order to derive evaporative mass-loss rates, we must specify the planetary structure. As our starting point, we assume planets with Earth-like solid cores, composed of 2/3 rock + 1/3 iron, swathed in a H/He envelope with a mass-fraction (defined as $M_{env}$/$M_{core}$) up to $\sim$1\%. As discussed above, this is consistent with both observations and theoretical calculations of the conditions at birth of ``Kepler'' planets \citep{wolfgang15b}. Current theories of planet formation, though, do not strongly constrain the initial entropy (or equivalently, radius) of such planets, with both ``hot start'' planets (those with short initial cooling time scales) and ``cold start'' ones (with long initial cooling time scales) remaining viable. We thus choose initial planetary radii (entropies) corresponding to cooling timescales in the range 10$^6$--10$^8$ yr, to comfortably span the plausible range from ``hot'' to ``cold start'' scenarios (further discussed in \S5). 
 
Finally, we are interested in the potential habitability, under the effects of evaporation, of low-mass planets in the HZ of M dwarfs. The ``classical'' HZ is defined as the range of orbital separations around a star where an Earth-mass planet with a CO$_2$--H$_2$O--N$_2$ atmosphere can sustain liquid water on the surface; the inner edge of this zone is (conservatively) set by the moist greenhouse effect, and the outer edge (again, conservatively) by the maximum greenhouse effect \citep{kasting93,kopparapuetal2013}. The position of the HZ of course changes as the star evolves in temperature and luminosity; what we are really interested in is the fate of planets located within the stable HZ of an M dwarf on the Main Sequence (MS). Using the fitting equations supplied by \citet{kopparapuetal2013} for the HZ boundaries, and the MS values of the stellar temperature and luminosity for a 0.4\,\msun\, dwarf from the \citet{baraffe98} tracks (\teff\, $\approx$ 3500\,K, \lbol\, $\approx$ 1.86$\times$10$^{-2}$\,\lsun; the star arrives on the MS after $\sim$500\,Myr), yields an inner edge of the classical HZ for our M dwarf at 0.15\,AU, and an outer edge at 0.28\,AU. For ease of calculation, and discussions for M dwarfs with masses around 0.4 M$_\odot$, we choose here to parametrise the HZ in terms of the blackbody temperature ($T_{\rm BB}$) of the planet instead, which we define as the equilibrium temperature of a planet with zero albedo, orbiting a star of luminosity \lbol\, at a radial separation of $a$: $T_{\rm BB} \equiv$ [\lbol/$(16\pi \sigma a^2)]^{1/4}$. We set the inner edge of the classical HZ around our M dwarf to be at a MS blackbody temperature of $\tbb1 = 300$\,K, and the outer edge at $\tbb1 = 200$\,K. These imply radii of $a$ = 0.12\,AU and 0.26\,AU respectively on the MS: very close to the actual separations from the  \cite{kopparapuetal2013} equations. Since we are interested in framing the general theory of evaporation driven habitability here, the $\sim$10--20\% difference has no appreciable effect on our conclusions. Detailed calculations of the habitability of systems with specific parameters (i.e. future detected exoplanet systems), can wait until a later date.  

We discuss our theory of X-ray driven evaporation, in both the hydrodynamic and ballistic limits, in \S3; our numerical implementation of this in \S4; and our treatment of thermal evolution in \S5. We present our results in \S6, and explore the implications for habitability around M dwarfs in \S7; our main conclusions are summarised in \S8.  

\section{Evaporation: Theory}

\subsection{X-ray Heating}
Our interest lies in low mass planets; in these, atmospheric heating and evaporation are dominated by X-rays \citep{oj12}. It is well known that, in radiative equilibrium, heating due to high energy photons can be described using a relationship between temperature ($T$) and ionization parameter ($\xi=F_x/4\pi n$, where $F_x$ is the received X-ray flux and $n$ the number density of particles) \citep{owen10}. \citet{oj12} used this principle to calculate  solutions to the evaporation problem. However, the precise form of the $T(\xi)$ relation is sensitive to the input X-ray spectrum and the metallicity of the gas. The X-ray spectrum of M dwarfs is quite different from that of solar type stars, being in general much softer in the $\sim 0.1-1$~keV range important for photo-electric heating. We must therefore calculate a new $T(\xi)$ relation, specific to our template M dwarf, to understand the behaviour of the X-ray irradiated planetary atmosphere. 

\begin{figure}
\centering
\includegraphics[width=\columnwidth]{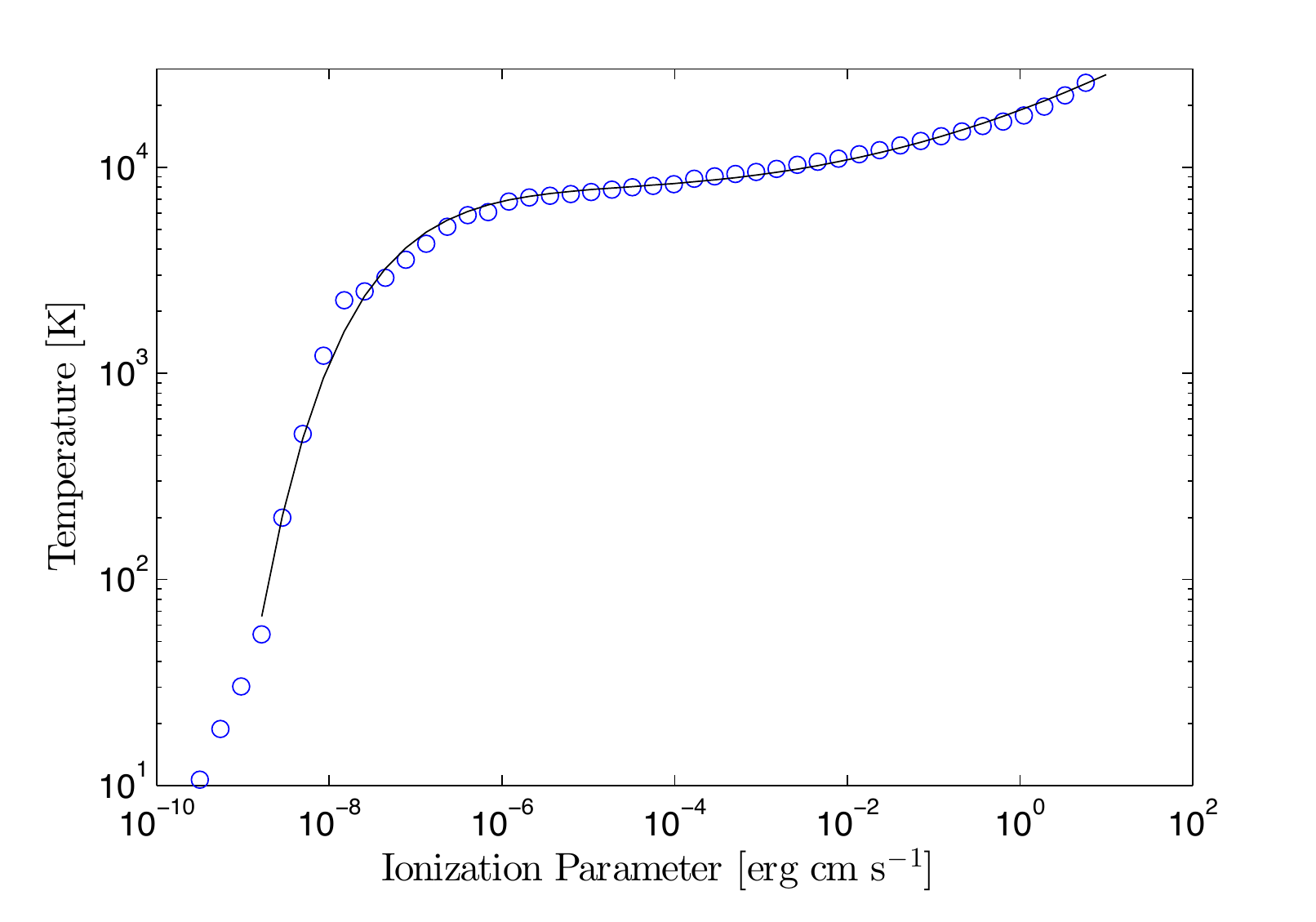}
\caption{Gas temperature as a function of ionization parameter from {\sc mocassin} radiative transfer calculations ({\it open circles}), along with a functional fit ({\it solid line}) used for our hydrodynamic calculations.}\label{fig:t-xi}
\end{figure}

We use the radiative transfer code {\sc mocassin} \citep{moc1,moc2,moc3} to solve for the $T(\xi)$ relation appropriate to the AD Leo X-ray spectrum discussed in \S2.2. We irradiate a plane parallel atmosphere, with densities spanning $10^{-8}-10^{-20}$ g cm$^{-3}$, the expected range in the evaporative flow. For specificity, we scale the X-ray flux to that expected at a planet orbiting at $\tbb1 = 300$\,K ($\sim$0.12\,AU); however, since we are concerned with mapping out the $T(\xi)$ profile, rather than with the X-ray flux itself, the exact flux chosen to perform the calculation does not matter very much. The resulting $T(\xi)$ profile is shown in Fig.\,\ref{fig:t-xi}, along with a functional fit which we adopt in our further calculations.

The soft M dwarf X-ray spectrum has a higher fraction of $\sim$0.1-1\,keV photons compared to solar-type spectra (see Fig.\,\ref{fig:spec}); consequently, the M dwarf produces considerably higher gas temperatures at low ionization parameters ($\xi < 10^{-5}$). We first attempt to solve for the evaporative flow by inserting our $T(\xi)$ relation into the semi-analytical methodology of \citet{oj12}, which assumes radiative equilibrium in the flow. This, however, leads to an inconsistency: over much of the parameter range of interest, the mechanical luminosity of the resulting evaporative wind becomes comparable to the radiative energy input rate, implying that radiative equilibrium is not a good approximation; instead, energy losses due to $P{\rm d}V$ work may dominate over radiative losses much of the time. In hindsight, this is not too surprising, given the higher heating efficiency of our M dwarf X-ray spectrum relative to that in the original \citet{oj12} calculations. As such, we must derive a full numerical solution to the radiation-hydrodynamic problem.  Nevertheless, the fact that, {\it when} radiative equilibrium holds, the gas temperature should equal that given by the $T(\xi)$ relation enables us to simplify the numerics, fully span the pertinent parameter space and, crucially, smoothly transition to the situation where radiative losses {\it do} become important (which, as discussed in \S2.1, ultimately determines when our hydrodynamic wind is quenched). Our specific numerical technique for accomplishing this is detailed in \S4.1.      

\subsection{Ballistic Mass-loss Estimate}\label{sec:jeans}

To assess the importance of {\it non}-hydrodynamic mass loss,  we calculate the ballistic (Jeans escape) mass-loss rate, given by
\begin{equation}
\dot{m}_{\rm JE}=4\pi R_p^2\rho_{\rm exo}\int_{v_{\rm esc}}^\infty\!\!\!\!{\rm d}v\,v f(v) \label{eqn:mdot_jeans}
\end{equation}
where $R_p$ is the planetary radius, $\rho_{\rm exo}$ the density at the exobase and $f(v)$ the distribution of particle velocities.  We assume that the exobase is in local thermodynamic equilibrium with the X-ray irradiation, so that its temperature is specified by our $T(\xi)$ relation; $f(v)$ is then the Maxwell-Boltzmann distribution at this temperature. The exobase density is
\begin{equation}
\rho_{\rm exo}=\frac{\mu}{H\sigma}
\end{equation}
where $H$ is the scale height of the atmosphere; $\mu$ the mean molecular weight of the gas particles, set to 1.35 for an atomic solar-abundance H/He mixture; and $\sigma$ the collision cross-section, for which we adopt $3.33\times10^{-15}$ cm$^{2}$ \citep{hunten73,tian05}. Equation~(\ref{eqn:mdot_jeans}) can be solved numerically for the mass-loss rate as a function of core mass. Since we expect the Jeans escape rates to be low, we assume the extent of the H/He atmosphere is small, and approximate the planetary radius $R_p$ by that of the solid core. The core mass-radius relation is from \citet{fortney07}, for a core comprising 2/3 rock and 1/3 iron by mass. 

\begin{figure}
\centering
\includegraphics[width=\columnwidth]{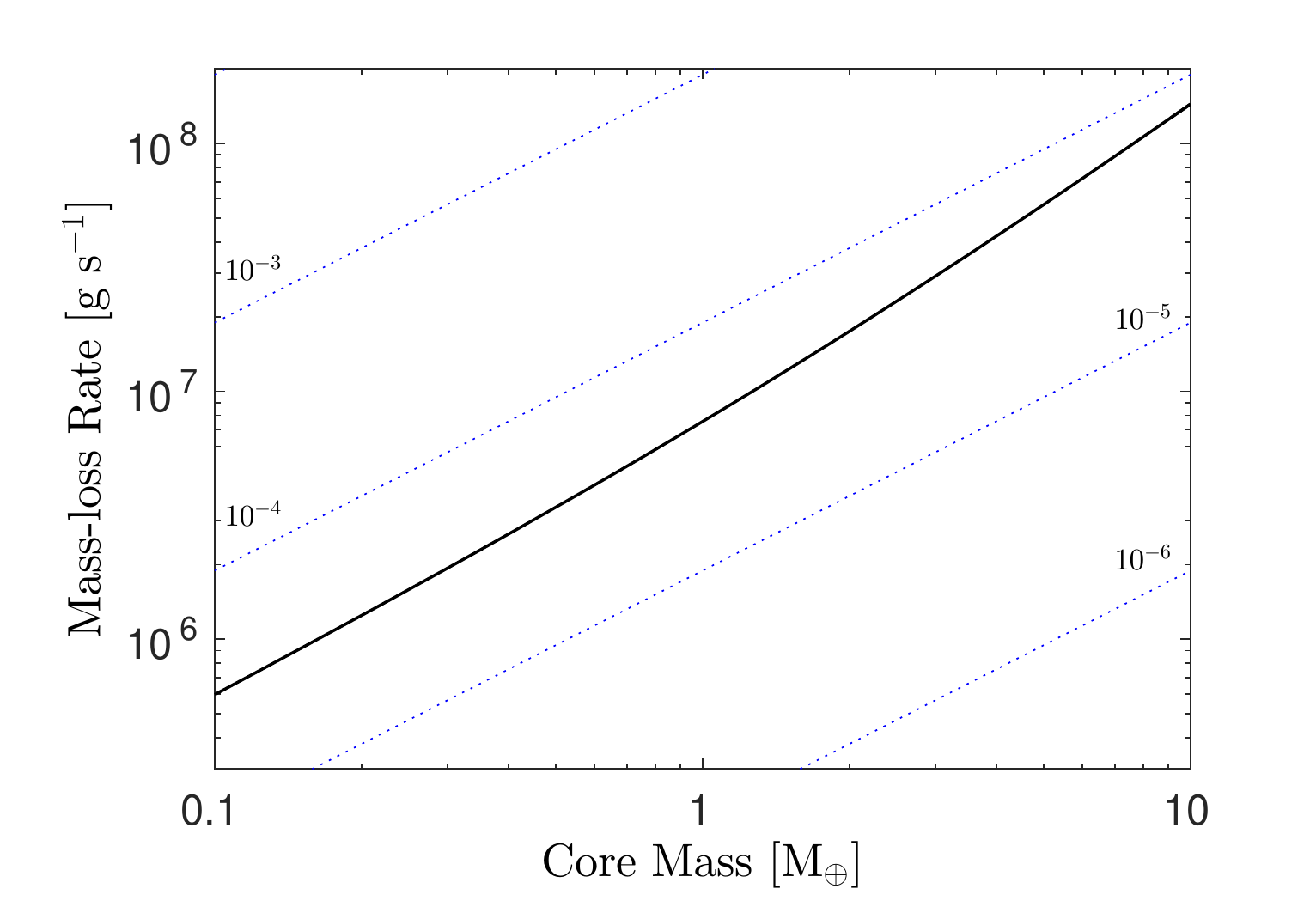}
\caption{Mass-loss rate due to Jean's escape, as a function of mass for a solid core made of 2/3 rock and 1/3 iron ({\it solid line}). The core mass-radius relation is taken from the fitting formulae provided by \citet{fortney07,fortney07_erratum}. The {\it dotted lines} show the required mass-loss rates to remove envelope mass-fractions of $10^{-3}$, $10^{-4}$, $10^{-5}$, $10^{-6}$ in 1~Gyr.}\label{fig:jeans}
\end{figure}

The resulting $\dot{m}_{\rm JE}$ are plotted in Fig \,\ref{fig:jeans}  as a function of core mass {\bc and are compatible with those found previously for highly irradiated terrestrial mass planets \citep[e.g.][]{tian08,erkaev13}}. These mass-loss rates are indeed very low; in particular, they are significantly below the rates one might expect from hydrodynamic evaporation (as we will see). We find that Jeans escape can only remove a H/He envelope mass-fraction ($M_{\rm env}/M_{\rm core}$) of order 10$^{-5}$ over a Gyr timescale: completely negligible compared to the $\sim$1\% initial H/He mass-fractions we are concerned with here. Thus, in order to strip any significant H/He envelope from solid terrestrial-mass cores in the HZ of an M dwarf, efficient hydrodynamic evaporation appears essential.

\section{Evaporation: Numerical Calculations}

As discussed in \S3.1, we must resort to full numerical hydrodynamic calculations to obtain the evaporation rates for the M dwarf case. Furthermore, to avoid the problems faced by \citet{mc09} in finding steady-state solutions to the wind problem, we explicitly integrate the time-dependent hydrodynamic equations until a steady-state is reached after several sound-crossing times \citep[e.g][]{oa15}: in all our calculations, we run the simulations for 30 crossing times to ensure that steady-state is achieved\footnote{This is confirmed by checking that a pseudo-Bernoulli function is conserved.}.

We employ the {\sc zeus} hydrodynamical code \citep{stone92,hayes06}, which has been successfully tested and used for planetary evaporation problems in the past \citep[e.g.][]{sp09,oa14,trammell14,oa15}. Our modified {\sc zeus} algorithm includes heating by X-rays and their radiative transfer \citep[c.f.][]{owen10,haworth16}.

\subsection{X-ray Heating and Radiative Transfer}

As noted in \S3.1, our evaporative flow may be in a state far from radiative equilibrium. Specifically, we expect the gas to behave as follows. In steady-state, since the flow is expanding and we do not anticipate any heating by shocks, the gas will always be cooler than the temperature at radiative equilibrium ($T_{\rm eq}$, given by the $T(\xi)$ relation). When the gas is considerably cooler than  $T_{\rm eq}$ - due to a large amount of $P{\rm d}V$ work - the photo-heating rate will greatly exceed the radiative cooling rate, and we may ignore radiative cooling. As the gas temperature approaches $T_{\rm eq}$, however, radiative cooling will become comparable to the photo-heating rate, and the temperature will asymptote to $T_{\rm eq}$. With this qualitative picture in mind, we build a simplified thermal model for X-ray heating / radiative cooling, using a ``Newtonian-heating'' approach. We begin by defining a {\it local} heating time as $t_{\rm th}\equiv \mathcal{E}_{\rm eq}/\dot{e}_{\rm abs}$, where $\mathcal{E}_{\rm eq}$ is the gas internal energy per unit mass at $T=T_{\rm eq}$, and $\dot{e}_{\rm abs}$ is the local heating rate per unit mass due to absorption of high energy photons, given by
\begin{equation}
\dot{e}_{\rm abs}(r)=\int\!\!\zeta_\nu F_\nu\exp[-\tau_\nu(r)]\frac{\sigma_\nu}{\mu}\, {\rm d}\nu
\end{equation}
where $\tau_\nu$ is the optical depth (from above) at a distance $r$ from the centre of the planet:
\begin{equation}
\tau_\nu(r)=\int_\infty^r\!\!n\sigma_\nu\, {\rm d}r .
\end{equation} 
Here, $\zeta_\nu$ represents the fraction of the photon energy that is converted into heating rather than used up in ionization. We caution that this is {\it not to be confused} with the efficiency parameter $\eta$ commonly deployed in ``energy-limited'' estimates of the mass-loss rate \citep[e.g.][]{lammer03,lammer09,erkaev07,lopez12}, which (arbitrarily) sets how much of the photon energy goes into $P{\rm d}V$ work instead of being radiatively lost. In our analysis, $\zeta_\nu$ merely accounts for the energy diverted to ionization processes; radiative energy losses, on the other hand, are explicitly accounted for by incorporating our derived $T(\xi)$ relationship into the calculations (via $\mathcal{E}_{\rm eq}$ in the energy equation, as described further below), and {\it not} via an ad hoc efficiency parameter $\eta$. 

Integrating a frequency-dependent radiation-hydrodynamics problem is beyond the scope of this work. Instead, we use the fact that the heating in our case is dominated by soft X-ray photons, mainly at energies of $\sim$1\,keV \citep{ercolano08}. We can thus approximate our frequency-dependent integral in Equation\,(3) by a monochromatic calculation at 1\,keV, with an absorption cross-section of $\sigma_{\rm 1keV}\approx10^{-22}$~cm$^{2}$. Heating by $\sim$1\,keV photons is primarily due to secondary photo-electrons; $\zeta_{\rm 1keV}$ then depends on the ionization state of the gas, with a value of $\sim$0.1 in neutral gas and $\sim$0.9 in highly ionized plasma \citep{xu91}. While we do not attempt to solve for the full ionization structure of the gas, {\bc  our {\sc mocassin} results indicate the gas is partially ionized \citep[e.g.][]{mc09,oa14,oa15}, so we pick an intermediate value of $\zeta_{\rm 1keV}=0.4$ consistent with the ionization fraction. Our calculations should not be exceptionally sensitive to our choice, as the thermostatting of the flow towards the $T(\xi)$ relationship will weaken the sensitivity of the flow to the choice of $\zeta$. }  

These considerations enable us to include radiative heating and cooling explicitly in our calculations, by writing the internal energy equation as
\begin{equation}
\frac{{\rm D}\mathcal{E}}{{\rm D}t} = -(\gamma-1)\mathcal{E}\nabla\cdot{\bf u} -\frac{\left(\mathcal{E}-\mathcal{E}_{\rm eq}\right)}{t_{\rm th}}\label{eqn:energy_evolve}
\end{equation}
where $\mathcal{E}$ is the specific internal energy of a fluid element and $\gamma$ is the ratio of specific heats, set to 5/3 for atomic gas. The first term on the R.H.S. of Equation\,(5) is the sink in internal energy due to $P{\rm d}V$ cooling, while the second is the sum of the source and sink terms due to radiative heating and cooling respectively. When $\mathcal{E}\ll \mathcal{E}_{\rm eq}$, this term is dominated by heating, with negligible cooling by radiation; when $\mathcal{E}\sim \mathcal{E}_{\rm eq}$, radiative heating and cooling rates begin to balance each other (they must be equal when $T=T_{\rm eq}$); and finally, when $t_{\rm th}\ll t_{\rm flow}$, this term dominates over $P{\rm d}V$ cooling and thermostats the gas temperature to $T=T_{\rm eq}$. 

Equation~(\ref{eqn:energy_evolve}) allows us to satisfy energy conservation, so that we do not overestimate the fraction of incoming X-ray flux that is converted to mechanical luminosity, while also accounting for the burgeoning amount of energy lost as radiation as the gas temperature draws closer to the radiative equilibrium value. This approach is much more satisfactory than the typical assumption of a constant ``heating efficiency'' everywhere \citep[e.g.][]{watson81,tian05,erkaev13,lammer14}, since it allows regions to locally account for radiative losses and thus varying the net heating efficiency in both a  local and global manner. Hence, in addition to a monochromatic ray-tracing calculation for 1\,keV photons, we solve Equation~(\ref{eqn:energy_evolve}) along with the time-dependent equations of hydrodynamics. We note that our model explicitly assumes that $T < T_{\rm eq}$, i.e., that we approach thermal equilibrium from below. This is satisfied here since the flow is expanding (and hence cooling via $P{\rm d}V$ work) everywhere. If this were not true -- e.g. if there was heating due to compression or shocks -- then Equation~(\ref{eqn:energy_evolve}) would need to be generalised to include a cooling timescale, in addition to the heating timescale $t_{\rm th}$. 

We solve our evaporation problem in 1D along a streamline connecting the star and the planet, within a spherical co-ordinate system (we thus implicitly assume that the velocity divergence is that of spherically symmetric outflow). We account for stellar gravity by using an effective gravitational potential of the form
\begin{equation}
\psi_{\rm eff}(r) = -\frac{GM_p}{r} -\frac{3GM_*r^2}{2a^3}
\end{equation}
where $M_p$ and $M_\ast$ are the planetary and stellar masses respectively, $r$ is the radial distance from the centre of the planet, and $a$ the orbital semi-major axis. We do not however include a contribution from the Coriolis force, since it is negligible in setting the mass-loss rates \citep{sp09,mc09,oj12,tripathi15}. 

\subsection{Hydrodynamic vs Ballistic Mass-loss}
While our code solves the hydrodynamic problem over the entire input parameter space, the applicability of the equations of hydrodynamics must be verified {\it a posteriori}, once the steady-state flow solution is obtained. Whether a flow is hydrodynamic or ballistic depends on whether the gas particles are collisional or not on a length-scale shorter than the flow-scale {\bc \citep[e.g.][]{johnson13}}. This property is characterised by the Knudsen number:
\begin{equation}
{\rm Kn}=\frac{\lambda}{\ell}
\end{equation}
where $\lambda\equiv1/n\sigma_{\rm col}$ is the mean free path of the gas particles. The relevant flow length-scale for the problem is the pressure scale-height, given by
\begin{equation}
\ell=\left(\frac{\partial\log P}{\partial r}\right)^{-1}.
\end{equation}
 Clearly, ${\rm Kn}$ increases with distance from the planet (since the particle number density $n$ decreases as an exponential function of the pressure scale-height; note that the exobase is the radius at which ${\rm Kn}=1$). At the same time, our hydrodynamic flow solutions are valid only if the gas remains in the hydrodynamic limit (${\rm Kn}\leq1$) all the way to the sonic surface; otherwise, the flow loses causal contact with the planet (and thus the ability to be influenced by upstream conditions) before the sonic point is reached (rendering the concept of a sonic point meaningless). Putting these two facts together, we follow \citet{oj12} in defining the evaporation to occur in the hydrodynamic regime if ${\rm Kn}\leq1$ at the sonic point.    

\subsection{Mass-loss Rates}

Using the framework described above, we construct a grid of evaporative mass-loss rates as a function of planetary mass, planetary radius and stellar X-ray luminosity, for a specified orbital distance. In order to comfortably cover the desired parameter space, we span planet masses ranging from 0.5 to 20\,\me; radii ranging from a value corresponding to a density of 10\,g\,cm$^{-3}$ (for a given planet mass) to half a Hill radius at the given orbital separation; and X-ray luminosities (evenly spaced logarithmically) ranging from 10$^{27}$ to 10$^{30}$\,erg\,s$^{-1}$. This grid is then coupled to the planetary thermal evolution code, to finally yield the detailed evolution of a given planet and its atmosphere. Before proceeding to a discussion of thermal evolution (\S5) and the final results for individual planets (\S6), however, it is instructive to examine a slice through our mass-loss grid for a fiducial orbit and stellar X-ray luminosity. 

\begin{figure}
\centering
\includegraphics[width=\columnwidth]{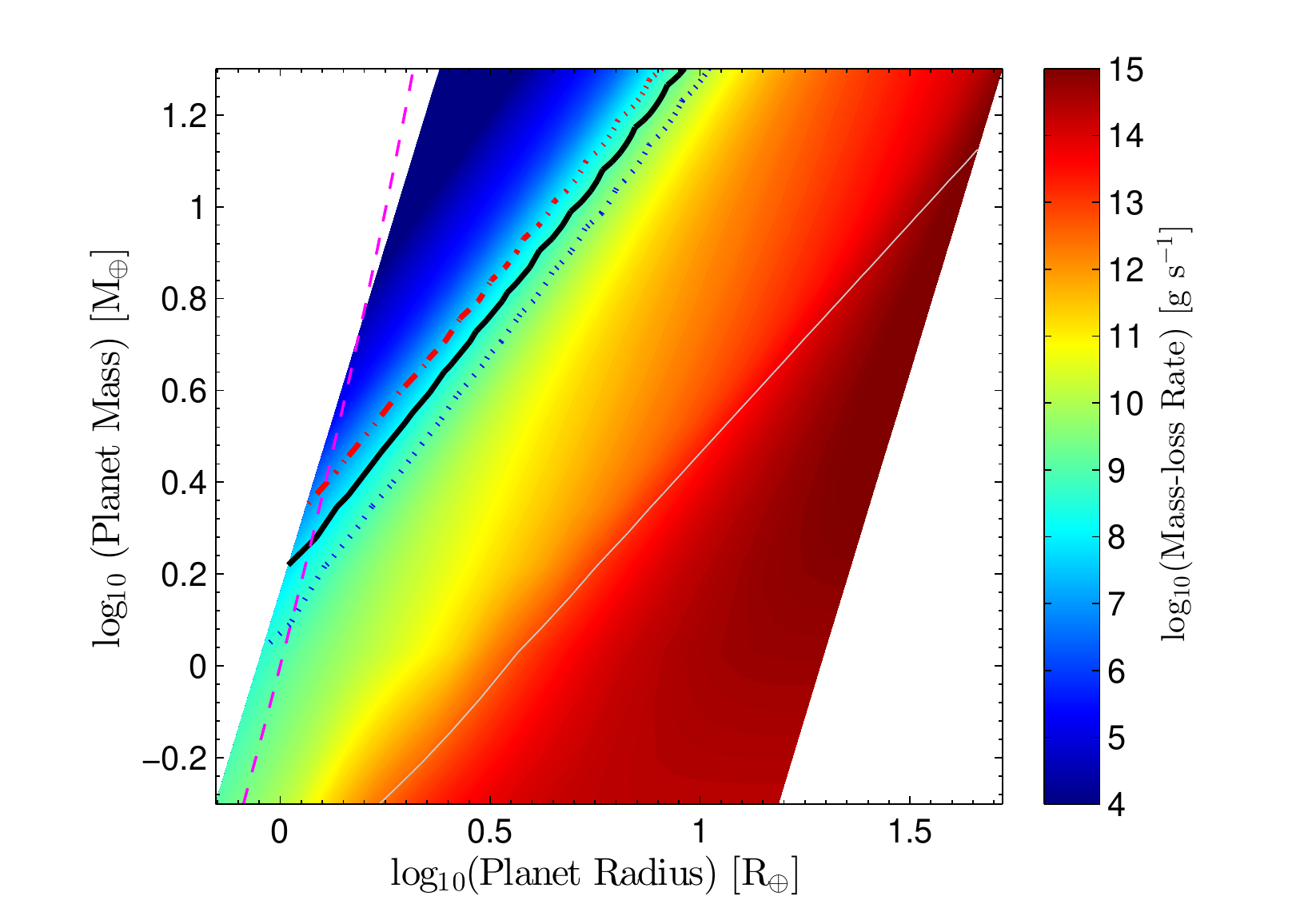}
\caption{Mass-loss rates as a function of planetary mass and radius, for an incident stellar X-ray flux of $1.23\times10^{4}$~erg~s$^{-1}$~cm$^{-2}$. The thick lines show contours of constant Knudsen number at the sonic point (Kn$_s$) of 10 ({\it red dot-dashed}), 1.0 ({\it black solid}) \& 0.1 ({\it blue dotted}). The {\it purple thin dashed line} shows the mass-radius relation for a solid core comprising 2/3 rock and 1/3 iron, taken from \citet{fortney07}. The {\it white thin solid line} shows a line of constant binding energy to thermal energy in the bolometrically heated atmosphere; planets to the right of this line have underlying envelopes too loosely bound to be considered tethered to the core (see text).}
\label{fig:mdot_rates}
\end{figure}

Fig.\,\ref{fig:mdot_rates} shows mass-loss rates as a function of planetary mass and radius, for an X-ray flux of 1.23$\times10^{4}$ erg~s$^{-1}$: the irradiation expected from AD Leo at $\sim$100\,Myr (roughly its current age) at an orbital separation of $a \approx 0.13$\,AU (corresponding to $\tbb1 \approx 300$\,K, i.e., the inner edge of the HZ at 1\,Gyr). Over-plotted is the mass-radius relationship for a solid core comprising 2/3 rock and 1/3 iron ({\it dashed magenta line}). A planet with a H/He envelope mass-fraction $\lesssim$1\%, for which the radius is dominated by the solid core, will lie only slightly to the right of this line ($\sim$ factor 2, \citealt{lopez14}), while a planet with a larger envelope mass-fraction, with radius dominated by the envelope, will lie much further to the right. 

Also over-plotted are lines of constant Knudsen number at the sonic point (which we will denote by Kn$_s$): Kn$_s$ = 10 ({\it red dot-dash line}), 1 ({\it solid black line}) and 0.1 ({\it blue dotted line}). For a fixed planetary mass, we see that Kn$_s$ decreases with increasing radius. Thus (recalling that hydrodynamic escape requires Kn$_s$ $\leq$ 1), it is easier to hydrodynamically evaporate a more distended atmosphere; an intuitive result, since more puffed-up atmospheres have a lower gravitational binding energy for a given planetary mass. Indeed, planets with radii to the right of the {thin white line} in the plot have atmospheres that are so loosely bound that they are unlikely to remain tethered to the planet for more than a Myr: our hydrodynamic calculations indicate that the evaporative flows on these planets are driven entirely by stellar bolometric heating rather than by X-ray heating, leading to a rapid ``boil-off'' phase during which most of the atmosphere is lost shortly after disc dispersal \citep[see][for a discussion of this process]{ow15}, see we are considering planets with relatively low envelope mass fractions this effect is only prominent in one case.      

Finally, we also see that the lines of constant Kn$_s$ are shallower than the mass-radius relationship for our solid core, with the Kn$_s$=1 line intersecting the latter locus at $M_p \sim 2$\,\me. Since hydrodynamic flows can only occur to the right of the Kn$_s$=1 line (because they require Kn$_s$ $\leq$ 1), only planets less massive than $\sim$2\,\me\, can be plausibly stripped entirely of their H/He envelopes via such flows. Higher mass planets can still undergo hydrodynamic evaporation as long as their envelopes remain very extended, but such flows will stall once the planetary radius shrinks sufficiently to hit the Kn$_s$=1 limit, which occurs while these planets still retain H/He mass-fractions $>$1\%. Subsequent mass loss only happens ballistically, which can hardly put a dent in such atmospheres over Gyr timescales (as demonstrated in \S3.2); thus, planets more massive than $\sim$2\,\me\, will retain substantial H/He envelopes over their lifetimes. In short, it is easier to completely strip a lower mass planet of its H/He envelope than a higher mass one: not surprising, since escape velocity increases with planetary mass. 

The threshold mass of $\sim$2\,\me\, identified here, separating the envelope retention / stripping outcomes, is of course quantitatively valid only for the specific X-ray flux adopted in this snapshot plot, and only in the absence of thermal evolution (which is not accounted for in this plot). The true threshold will depend on the integrated history of the X-ray irradiation (since a higher X-ray flux, expected at earlier times, will push the sonic point deeper into the atmosphere, raising the threshold mass), as well as on the thermal evolution of the planet (since the atmosphere, contracting as it cools, sinks ever deeper into the planet's gravity well, making hydrodynamic escape harder and decreasing the threshold mass). By coupling our full mass-loss grid (of which Fig.\,\ref{fig:mdot_rates} is only an instantaneous slice at one X-ray flux) to our thermal evolution code, we will account for both of these effects in \S5 and \S6 below. Nevertheless, the physical effects summarised here will remain qualitatively valid, and only particularly low-mass cores can be stripped entirely by evaporation.

\section{Thermal Evolution}
In order to fully model the evolution in evaporative mass loss, we must account for how the H/He envelope cools and contracts. We solve the coupled thermal evolution and evaporation problem for a planet with a solid core and H/He envelope using the {\sc mesa} stellar evolution code adapted to planets \citep{paxton11,paxton13}. The method is described at length in \citet{ow13}; here we only summarise the salient inputs to the code.

\begin{figure}
\centering
\includegraphics[width=\columnwidth]{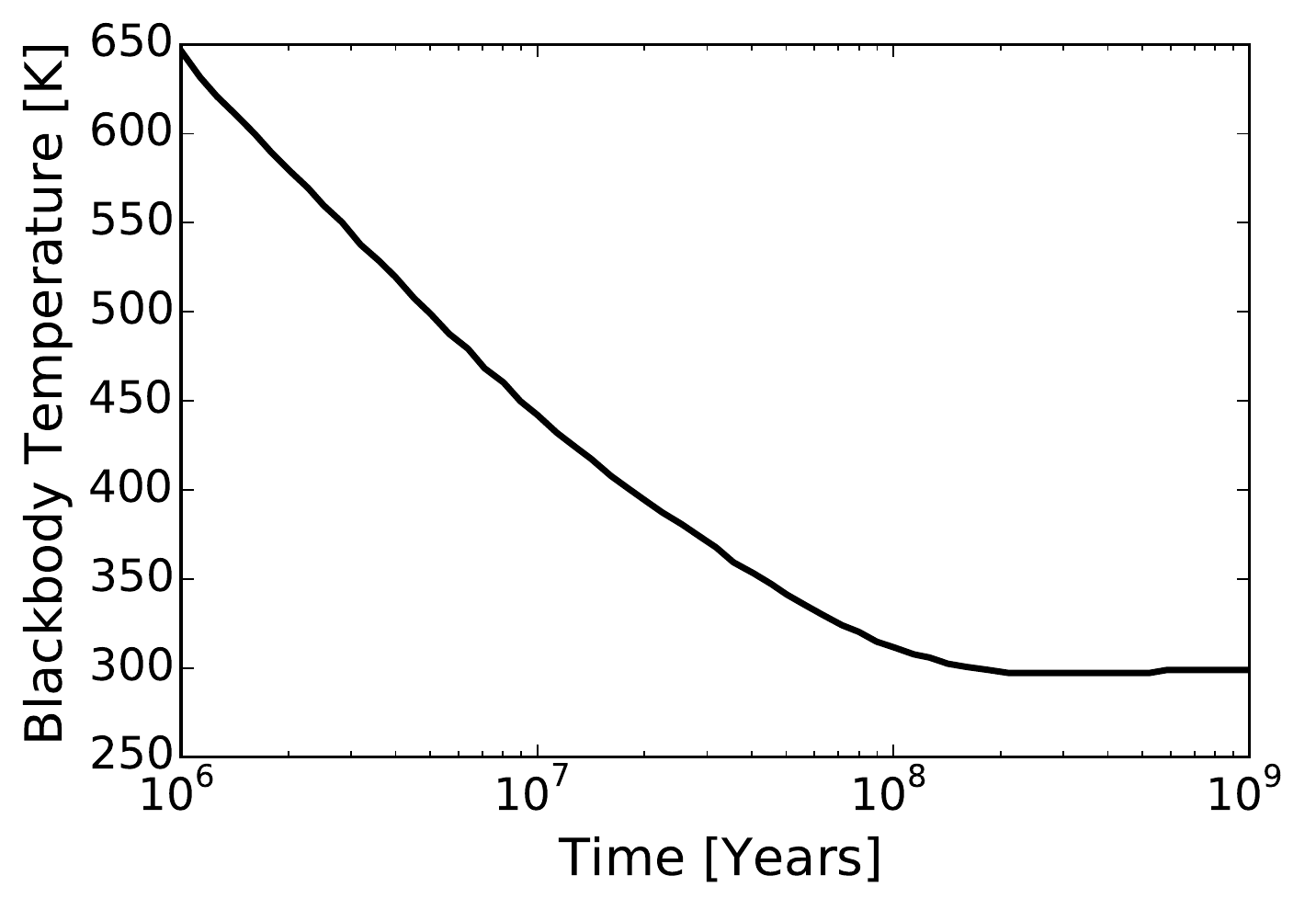}
\caption{Evolution of the blackbody temperature ($T_{\rm BB}$) for a planet at $\sim 0.12$~AU around an M-star.}
\label{fig:Teq_time}
\end{figure}

As discussed in \S2.2, the stellar bolometric luminosity ($L_{\ast}$) is assumed to follow the Lyon evolutionary track \citep{baraffe98} for a 0.4\,M$_{\odot}$ star (appropriate for AD Leo), while the planet's orbital distance is fixed at either $a \approx 0.12$\,AU or 0.26\,AU, so that the blackbody temperature at the planet's location at 1\,Gyr is $\tbb1 = 300$\,K or 200\,K respectively (corresponding to the inner and outer edges of the classical HZ around an M dwarf; see \S2.2). For reference, the evolution in blackbody temperature at 0.12\,AU is shown in Fig.\,\ref{fig:Teq_time}. 

The radius of the solid core is assumed to be constant throughout the planet's evolution, and is specified, for a given core mass and our adopted core composition of 2/3 rock + 1/3 iron, by the mass-radius relationship from \citet{fortney07,fortney07_erratum}. We do account for the thermal content of the core, due to both radioactive decay and thermal heat capacity, by adopting an Earth-like value \citep[see][]{lopez12}. The effects of both external bolometric irradiation by the star and internal release of thermal energy from the core are accounted for in {\sc mesa}, using the $T(\tau)$ relationship formulated by \citet{guillot10} for a gray atmospheric model with both internal and external heat fluxes, where one uses an incoming stream with a frequency set to the peak of the black-body spectrum for the star and an outgoing IR stream in local thermodynamic equilibrium with the upper atmosphere.  

Current planet formation models are unable to stringently constrain the initial thermal properties of newborn planets. Consequently, we consider planets with a large range of initial radii (or, more accurately, initial entropies). The latter are parametrised in terms of an initial cooling (i.e., Kelvin-Helmholz) timescale $\tcooli$, defined as the ratio of a planet's initial internal energy to its initial luminosity. Naively, one expects this Kelvin-Helmholz timescale to be of order the ``age'' of the planet at birth, i.e., the time it takes to form, and thus comparable to the mean lifetime of primordial disks, $\sim$10$^7$ yr \citep[e.g.][]{lee14}. A ten-fold uncertainty in this estimate is not infeasible, due to variations in disc lifetimes \citep[e.g.][]{hernandez07,ercolano11,owen11} or post-formation processes \citep[e.g.][]{ow15,liu15}.; as such, we use starting models with $\tcooli$ in the range 10$^6$--10$^8$ yr, to encompass a plausible maximal difference between ``hot start'' planets (with short initial cooling times) and ``cold start'' ones (with long initial cooling times) \citep[c.f.][]{marley07,spiegel12}. 

Finally, evaporation is included in our {\sc mesa} calculations by coupling the code to our grid of newly calculated mass-loss rates as a function of planetary mass, radius and stellar X-ray flux (\S4.3).

\section{Planetary Evolution: Results}

We present here our results for mass loss (accounting for both evaporation and thermal evolution) in the HZ of our fiducial M dwarf, for three core masses -- $\mcore$ = 0.8, 0.9 and 1\,\me\, -- at two different orbital separations: $a \approx 0.12$\,AU (roughly the inner edge of the HZ) and 0.26\,AU (roughly the outer edge). We investigate initial H/He mass-fractions (denoted henceforth by $\mfraci$) of $\sim$1\% and $\sim$0.04\% (i.e., $\mfraci \sim 10^{-2}$ and 4$\times$10$^{-4}$). We stop our calculations either after 1\,Gyr of evolution, or once the envelope mass-fraction falls to $10^{-5}$, the amount that can be efficiently removed by Jeans escape on less than Gyr timescales (see \S\ref{sec:jeans}). 

\subsection{Evolution Without Evaporation} 
We start by considering the evolution of the atmospheric temperature, density and pressure at the planetary surface ($\ts$, $\rhos$ and $\ps$ respectively), over Gyr timescales, in the {\it absence} of evaporation (i.e., with evaporation artificially turned off). These results constitute a baseline from which to judge whether evaporation will aid habitability or not. 

We first carry out this analysis for $\tbb1$ = 300\,K (i.e., at $a \sim 0.12$\,AU). Figs.\,6, 7 and 8 show our results for $\mcore$ = 0.8, 0.9 and 1.0\,\me\, respectively, for a H/He envelope with $\mfraci = 10^{-2}$. With evaporation turned off, of course, this mass fraction remains constant over time. In each case, we present results for both a relatively long initial cooling timescale ($\tcooli$) for the planetary atmosphere, corresponding to a ``cold start'' planet formation scenario, and a relatively short $\tcooli$, corresponding to a ``hot start'' (see \S5).   

We see that, for all three core masses, the final surface temperature after a Gyr is $\ts \sim 1000$\,K, and the final pressure $\ps$ is nearly 10$^4$ bar (with variations in $\tcooli$ making negligible difference to these results). These are well above the critical point of water ($T_{crit} \approx 647$\,K, $P_{crit} \approx 221$\, bar); as such, any surface water would exist as a supercritical fluid, not as a liquid (see phase diagram of water, Fig.\,20; constructed from phase transition equations provided by Martin Chaplin (pvt.\,comm.\,2015)\footnote{More detailed phase diagram for water by M.\,Chaplin available online at:\\ \url{http://www1.lsbu.ac.uk/water/water\_phase\_diagram.html\#bd}. The equations used for the phase boundaries can be seen by hovering over the boundaries in the latter plot.}). Hence, as stated earlier, removal of a large fraction of the H/He envelope via hydrodynamic escape, which will reduce $\ts$ and $\ps$, is {\it required} for habitability when $\mfraci$ is of order a percent. The degree to which this can happen is the focus of our analysis in \S\S6.1 and 6.2 below, since, as discussed, such relatively high initial mass fractions are commensurate with current data.      

Next, in Figs.\,9, 10 and 11, we plot the evolution of surface conditions for a much lower initial atmospheric mass: $\mfraci$ $\approx$ 4$\times10^{-4}$. Again, in the absence of evaporation, this fraction remains constant. Now the surface conditions after a Gyr are much more benign, with $\ts \sim 400$\,K and $\ps \sim 300$\,bar. As the phase diagram in Fig.\,20 shows, surface water should exist as a liquid under these conditions.  However, realistic levels of evaporation will in fact remove most or all of this tenuous atmosphere: as we will show below, strong hydrodynamic escape very quickly reduces the atmospheric mass-fraction to $\sim$10$^{-5}$; even Jeans escape can then remove the remainder on Gyr timescales. 

Any potential long-term habitability will then depend on the quantity, properties and evolution of (outgassed) secondary atmospheres, similar to the case on Earth {\bc and Mars \citep[e.g.][]{elkins08a}}. As such, for these low initial atmospheric mass fractions, we will plot the evolution of the mass fraction down to 10$^{-5}$, to make the point that the primordial H/He atmosphere is likely to be entirely lost, but will abstain from any further discussion of the final surface conditions, which will depend on secondary atmospheres whose modeling is beyond the scope of this paper \citep[{\rc although several authors have begun to investigate this stage theoretically; e.g.,}][]{elkins08b,elkins11}. Nonetheless, since secondary atmospheres are at least {\it present} on all the terrestrial planets and massive moons in the solar system, we will consider planets whose H/He envelopes have been completely stripped to still be {\bc ``potentially habitable''}. 

The results for $\tbb1 = 200$\,K ($a \sim 0.26$\,AU) without evaporation (not plotted) are similar: while the surface temperatures after a Gyr are now somewhat lower, $\ts$ and $\ps$ are still far too high for liquid surface water when $\mfraci = 10^{-2}$; for $\mfraci \approx 4\times10^{-4}$, the surface conditions are again conducive to liquid water, but with the same caveats as above. 

\subsection{Evaporation at Inner Edge of HZ ($\boldsymbol{\tbb1 = 300}$\,K)}
We now turn the evaporation on, and once again examine the results for core masses of 0.8, 0.9 and 1.0\,\me\, at $\tbb1 = 300$\,K. 

\noindent {\bf 0.8\,\me :} Fig.\,12 shows the evolution of the atmospheric mass-fraction and planetary radius for $M_{core}$ = 0.8\,\me, for $\mfraci \approx 10^{-2}$ (top panels) and 4$\times10^{-4}$ (bottom panels). In both cases, evaporation strips the atmosphere down to a mass fraction of 10$^{-5}$ extremely rapidly, in $\lesssim$10\,Myr. This steep decline is due to runaway evaporation, wherein the cooling timescale exceeds the evaporation timescale: the atmosphere cannot contract quickly enough to avoid evaporation by falling deeper into the planet's gravitational potential well \citep[c.f.][]{baraffe05}. Note that the runaway occurs much earlier than the median age of 10$^{8\pm0.5}$\,Myr estimated for AD Leo (and for other early to mid-M field dwarfs evincing similar saturated X-ray activity; see \S2.2); consequently, this evaporation result is unaffected by our simplistic assumption that activity remains saturated beyond a few 100 Myr up to a Gyr (instead of declining at later times).  

For a 0.8\,\me\, core, therefore, we expect any primordial H/He atmosphere with initial mass fraction $\lesssim$1\% to be completely lost: eroded down to 10$^{-5}$ in $\lesssim$10\,Myr by runaway hydrodynamic evaporation, as shown, and the minuscule remainder efficiently peeled off by continuing hydrodynamic and/or Jeans escape over a less than Gyr timescale.    

\noindent {\bf 0.9\,\me :} Fig.\,13 shows the evolution in atmospheric mass fraction and planetary radius for  $M_{core}$ = 0.9\,\me. For $\mfraci \approx 10^{-2}$, we see that the ``hot'' and ``cold'' start cases behave very differently. The atmosphere in the former is initially relatively bloated, resulting in a large fraction of it being blown off almost instantaneously when evaporation is switched on (a phenomenon termed ``boil-off'' by Owen \& Wu 2015; see \S4.3). For reasons of clarity, this initial boil-off, which occurs over the first few time-steps of our calculations (in $\lesssim$1.5 Myr), is masked out in Fig.\,13 and subsequent figures. Essentially one cannot build stable models with envelope mass fractions and cooling times as short as desired -- they are not globally thermodynamically stable in the presence of mass-loss. Specifically, in the pressure confining environment of the parent disc the planet can acquire a $\sim 1$\% envelope, but once this pressure confinement is removed such a massive envelope cannot be gravitationally bound to the planet (see Owen \& Wu, 2015). Therefore, the bolometrically driven mass-loss allows the planet to readjust to a stable state that is in quasi thermodynamic equilibrium. The plots trace the evolution of the planet after boil-off ends. The remainder then undergoes runaway hydrodynamic escape, reducing the mass fraction to 10$^{-5}$ in $\sim$10\,Myr. For a ``cold start'', on the other hand, the initial atmosphere is much more compact, and the pace of hydrodynamic evaporation is thus far more leisurely; by $\sim$30\,Myr, evaporation is too slight to make any significant difference, and the atmospheric mass fraction settles down at $\sim$10$^{-3}$. Jeans escape, which only becomes efficient at mass fractions $\lesssim$10$^{-5}$, has no discernible effect on the surviving atmosphere. Note as well that the $\sim$30\,Myr it takes for the mass fraction to reach approximate steady-state is at the lower limit of the 10$^{8\pm0.5}$\,Myr age estimate for AD Leo; our result is thus minimally affected by our naive assumption that activity continues to be saturated beyond a few 100 Myr instead of declining at such late times. Finally, cores with $\mfraci \approx 4\times$10$^{-4}$ are evaporated down to 10$^{-5}$ in only a few Myr, regardless of initial cooling timescale, just as in the 0.8\,\me\, case.      

The evolution of surface conditions for the $\mfraci \approx 10^{-2}$ case is shown in Fig.\,14. For a ``cold start'' 0.9\,\me\, core, we see that $\ts$ and $\ps$ after a Gyr are $\sim$ 530\,K and 630\,bar respectively (note that the slight continuing evolution of these quantities beyond $\sim$100\,Myr is predominantly due to ongoing thermal evolution of the planet, and not any appreciable evaporative mass loss). This [$\ts$, $\ps$] combination is conducive to liquid surface water (Fig.\,20), and within the range found in deep-ocean hydrothermals vents on Earth.     

Fig.\,14 also shows the ``hot start'' core with the same $\mfraci \approx 10^{-2}$ for comparison. The [$\ts$,$\ps$] $\sim$ [350\,K, 10\,bar] achieved at the end of our calculations is of no particular significance, since we expect the tiny remaining atmospheric mass fraction of 10$^{-5}$ to be completely removed subsequently by even Jeans escape; we only note that the surface temperature is beginning to flatten out since it cannot fall below the equilibrium temperature (with albedo assumed to be zero) of 300\,K at this orbital radius, while the pressure continues to plummet, which will eventually lead to any liquid surface water boiling into vapour. Of more physical interest is the fact that $\ts$ in the ``hot start'' case is {\it lower} than in the ``cold start'' one, once the rapid initial boil-off stage (which is not plotted in these figures) has ended. This is because hydrodynamic escape requires mass continuity at the sonic point; when a substantial fraction of the atmosphere is blown off from the top, therefore, the underlying layers swiftly expand to take its place, and the upward advection of heat in this process causes the surface to cool \citep[see discussion in][]{ow15}.        

In summary, at $\tbb1 = 300$\,K, we expect liquid water to survive on Gyr timescales on the surface of a ``cold start'' 0.9\,\me\, core with $\mfraci \approx 10^{-2}$; the partial evaporation of the H/He atmosphere engenders habitable conditions in this case. A ``hot start'' 0.9\,\me\, core with the same $\mfraci$, or a core with much smaller $\mfraci$ and either a ``hot'' or a ``cold'' start, however, is unlikely to retain any of its primordial H/He atmosphere; habitability conditions in this case will depend on any secondary atmosphere that arises. Therefore, a 0.9~M$_\oplus$ core appears to be at the {\it transition} between a planet becoming habitable due to evaporation, and remaining uninhabitable due to retention of a significant H/He envelope.    

\noindent {\bf 1.0\,\me :} Finally, Fig.\,15 shows the the evolution in atmospheric mass fraction and planetary radius for $M_{core}$ = 1.0\,\me. There is relatively little evaporation when $\mfraci \approx 10^{-2}$, with both ``hot'' and ``cold start'' cores equilibrating at a mass fraction of $\sim$2$\times$10$^{-3}$ (there is some initial boil-off in the ``hot start'' case, but hardly enough to initiate runaway evaporation; the planet's gravity is too strong). When $\mfraci \approx 4\times$10$^{-4}$, on the other hand, runaway evaporation reduces the fraction to 10$^{-5}$ in just a few Myr, as in the 0.8 and 0.9\,\me\, cases. 

In Fig.\,16 we plot the evolution of surface conditions for $\mfraci \approx 10^{-2}$. Both ``hot'' and ``cold'' start cores end up with very similar surface temperatures and pressures after a Gyr, with [$\ts$, $\ps$] $\sim$ [800\,K, 2$\times$10$^{3}$\,bar]. Fig.\,20 demonstrates that these conditions are too extreme for liquid water; any surface water here can only exist as a supercritical fluid.  

\subsection{Evaporation at Outer Edge of HZ ($\boldsymbol{\tbb1 = 200}$\,K)}
We now investigate the effects of hydrodynamic evaporation at $\tbb1  = 200$\,K. The single case of a 0.8\,\me\, core will serve to illustrate the main results here. 

\noindent {\bf 0.8\,\me :} We plot the evolution in atmospheric mass fraction and planetary radius for this core mass in Fig.\,17. There is hardly any evaporation for either $\mfraci \approx 10^{-2}$ or 4$\times10^{-4}$: in both cases, the final mass fraction after a Gyr is very similar to the initial value. Figs.\,18 and 19 show the evolution in surface conditions for these two initial mass fractions. For $\mfraci$ of 10$^{-2}$, the final [$\ts$, $\ps$] after a Gyr is $\sim$ [800\,K, 4$\times$10$^{3}$\,bar]; as the phase diagram in Fig.\,20 indicates, surface water will exist as a supercritical fluid instead of a liquid under these conditions. For $\mfraci$ of 4$\times$10${-4}$, on the other hand, the equilibrium [$\ts$, $\ps$] after a Gyr is $\sim$ [400\,K, 280\,bar], which {\it is} conducive to liquid water. 

We can easily extrapolate from these results, and the preceding ones, to deduce the implications for 0.9 and 1.0\,\me\, cores at this radial separation. First, surface temperature and pressure always increase with core mass, for a given $\mfraci$ and orbital radius. Thus, since a 0.8\,\me\, core with an initial H/He atmospheric mass fraction of a percent cannot harbour liquid water at $\tbb1 = 200$\,K, higher mass cores with the same initial mass fraction cannot either. Second, in the {\it absence} of evaporation, $\ts$ and $\ps$ are very similar for 0.8--1.0\,\me\, cores at any fixed $\mfraci$ and orbital radius (e.g., see the no-evaporation cases plotted in Fig.\,20 for $\tbb1 = 300$\,K: the $\ps$ for 0.8--1.0\,\me\, cores are nearly identical, and their $\ts$ vary by $<$100\,K, for both $\mfraci$ of 10$^{-2}$ and 4$\times$10$^{-4}$). We found above that there is hardly any evaporation in the 0.8\,\me\, case even with $\mfraci \approx 4\times$10$^{-4}$ at this orbital radius; therefore, since the evaporative rate decreases with increasing core mass (as the gravitational potential well deepens), we can invoke the no-evaporation results here for the entire range 0.8--1.0\,\me. This in turn implies that 0.9 and 1.0\,\me\, cores with $\mfraci \approx 4\times$10$^{-4}$ will have [$\ts$, $\ps$] very close to that of the 0.8\,\me\, core with the same initial mass fraction, and can thus also harbour liquid surface water on Gyr timescales. These inferences are indeed what our more detailed calculations show.   

\begin{figure}
\includegraphics[width=\columnwidth]{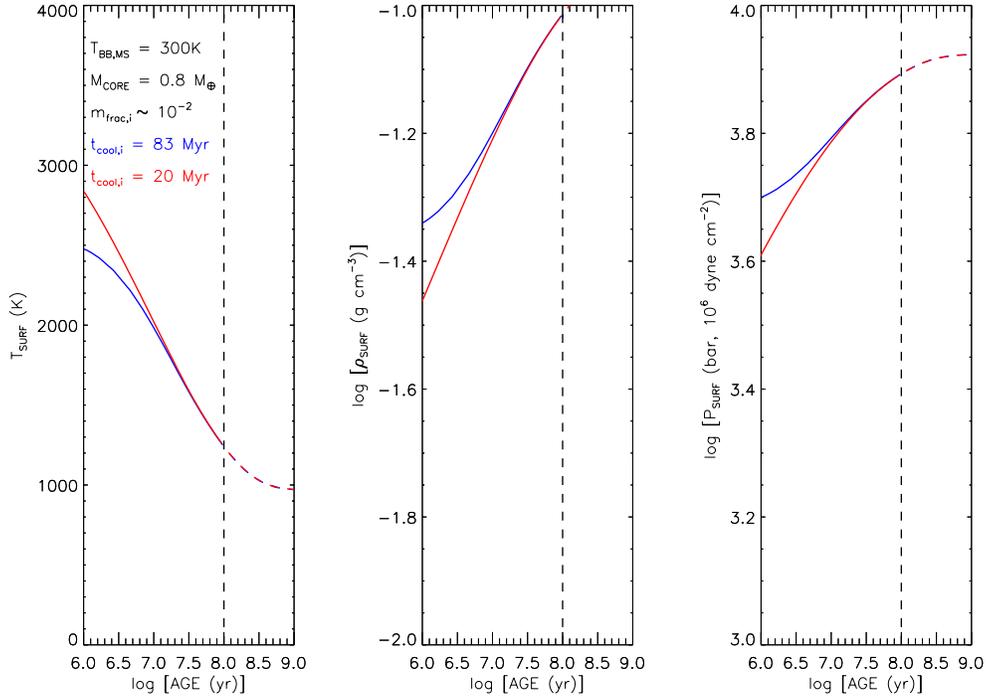}
\caption{Evolution of the surface values (i.e., at the envelope-core boundary) of the temperature (left panel), density (middle) and pressure (right) as a function of time, for planets evolving {\it without} the influence of evaporation. The planet is located at an orbital separation corresponding to $\tbb1$\, = 300\,K (inner edge of HZ), with a core mass $\mcore$\, = 0.8\,M$_{\oplus}$ and an initial H/He envelope mass-fraction of $m_{\rm frac,i}$ $\approx$ 10$^{-2}$ (the latter remains constant with time in these `no-evaporation' calculations). The {\it dashed vertical line} shows the median age of AD Leo ($\sim$100\,Myr). The {\it red curve} is for a ``hot-start'' model, and the {\it blue curve} for a ``cold-start'' model; $\tcooli$\, for each is noted in the left panel.}\label{fig:0.8ne_300_0.01}
\end{figure}

\begin{figure}
\centering
\includegraphics[width=\columnwidth]{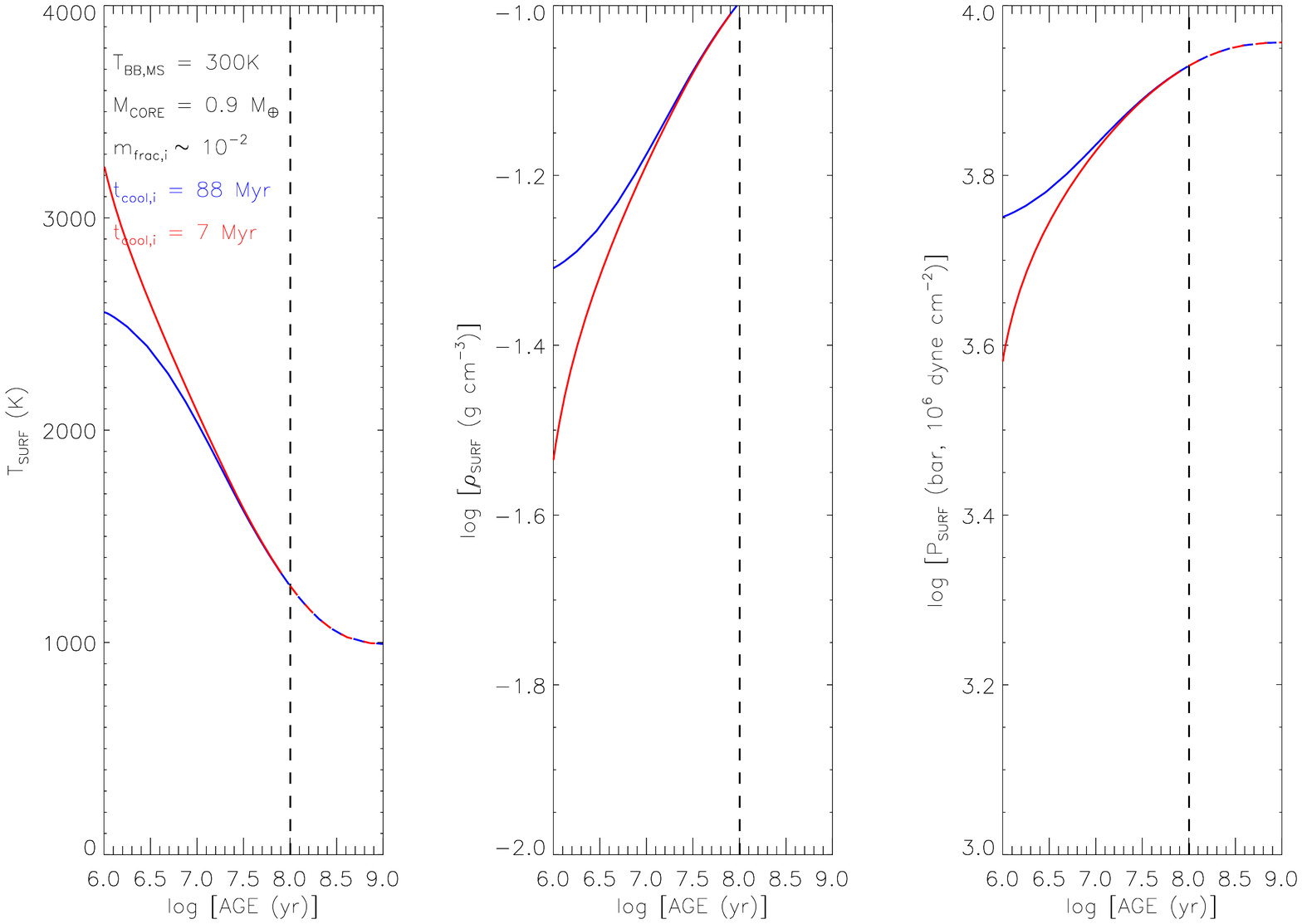}
\caption{Same `no-evaporation' calculation as in Figures~\ref{fig:0.8ne_300_0.01}, but for $\mcore$\, = 0.9\,M$_{\oplus}$.}\label{fig:0.9ne_300_0.01}
\end{figure}

\begin{figure}
\centering
\includegraphics[width=\columnwidth]{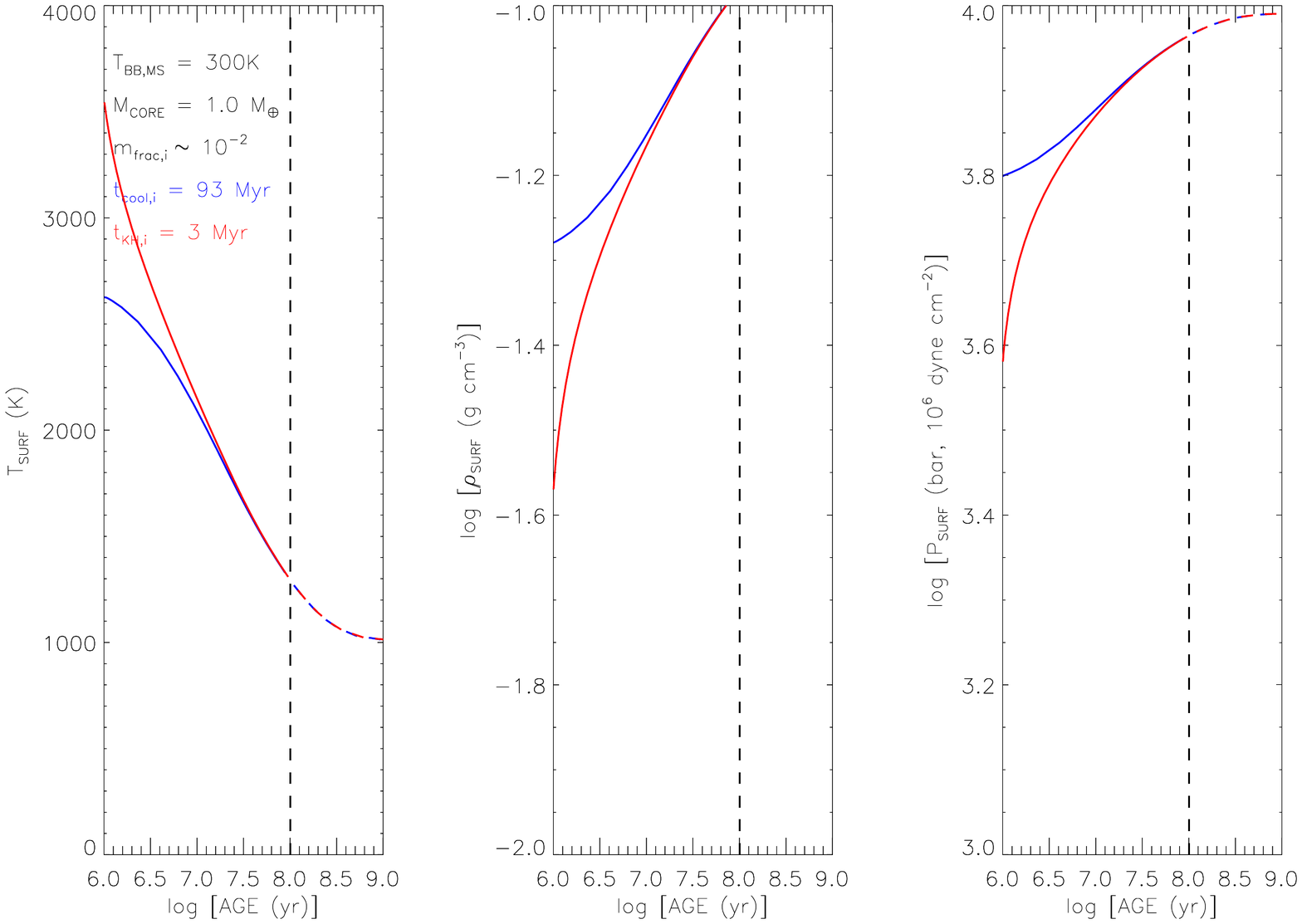}
\caption{Same `no-evaporation' calculation as in Figures~\ref{fig:0.8ne_300_0.01}, but for $\mcore$\, = 1.0\,M$_{\oplus}$.}
\label{fig:1.0ne_300_0.01}
\end{figure}

\begin{figure}
\centering
\includegraphics[width=\columnwidth]{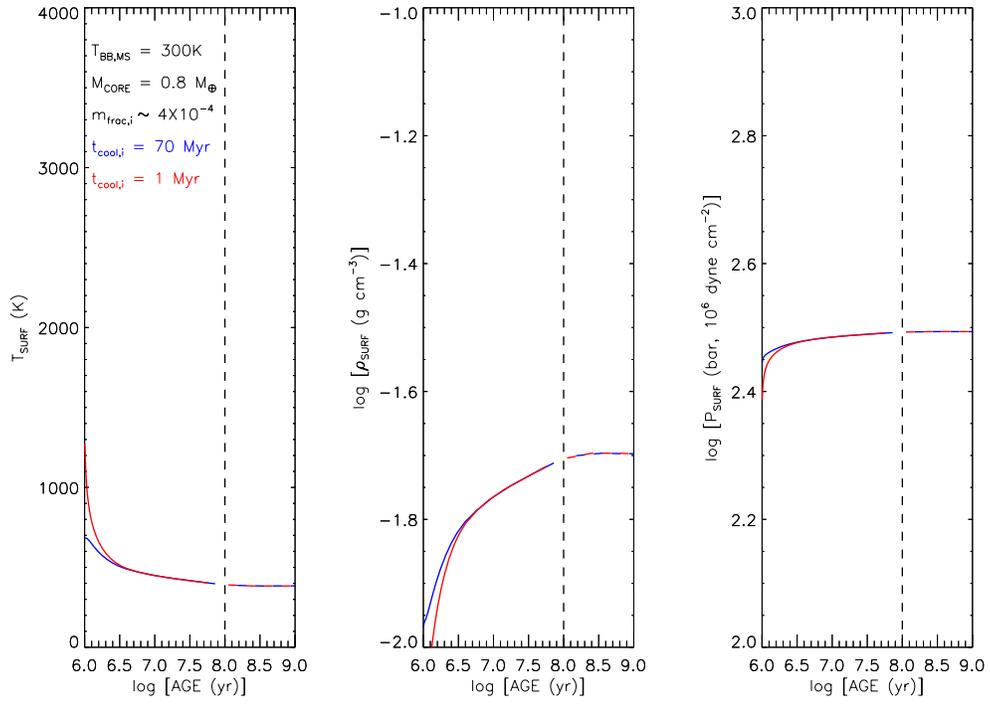}
\caption{Same `no-evaporation' calculation as in Figures~\ref{fig:0.8ne_300_0.01}, for the same core mass $\mcore$\, = 0.8\,M$_{\oplus}$, but now for an initial H/He envelope mass-fraction of $m_{\rm frac,i}$ $\approx$ 4$\times$10$^{-4}$.}\label{fig:0.8ne_300_4e-4}
\end{figure}

\begin{figure}
\centering
\includegraphics[width=\columnwidth]{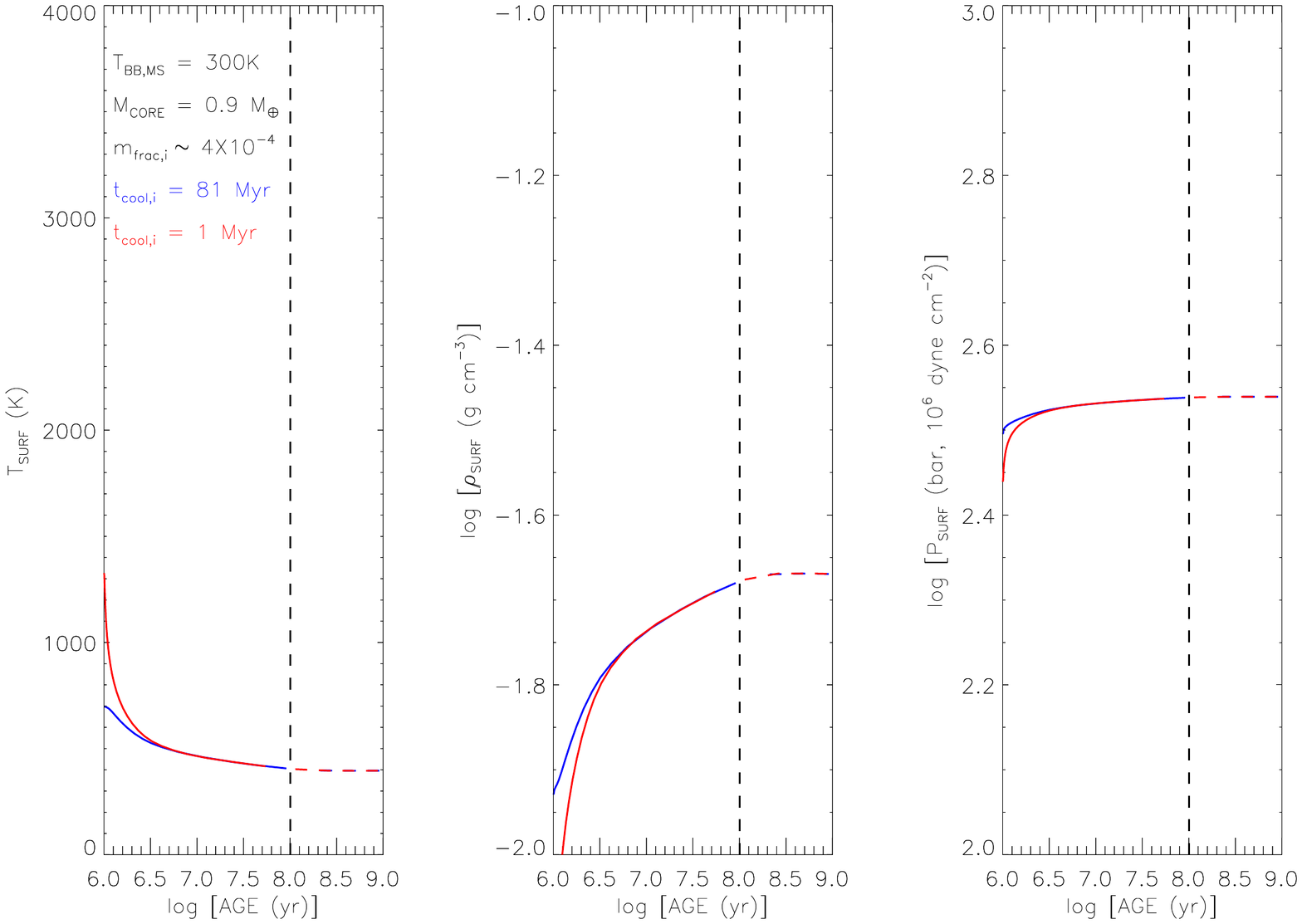}
\caption{Same `no-evaporation' calculation as in Figures~\ref{fig:0.8ne_300_4e-4}, but for $\mcore$\, = 0.9\,M$_{\oplus}$.}\label{fig:0.9ne_300_4e-4}
\end{figure}

\begin{figure}
\centering
\includegraphics[width=\columnwidth]{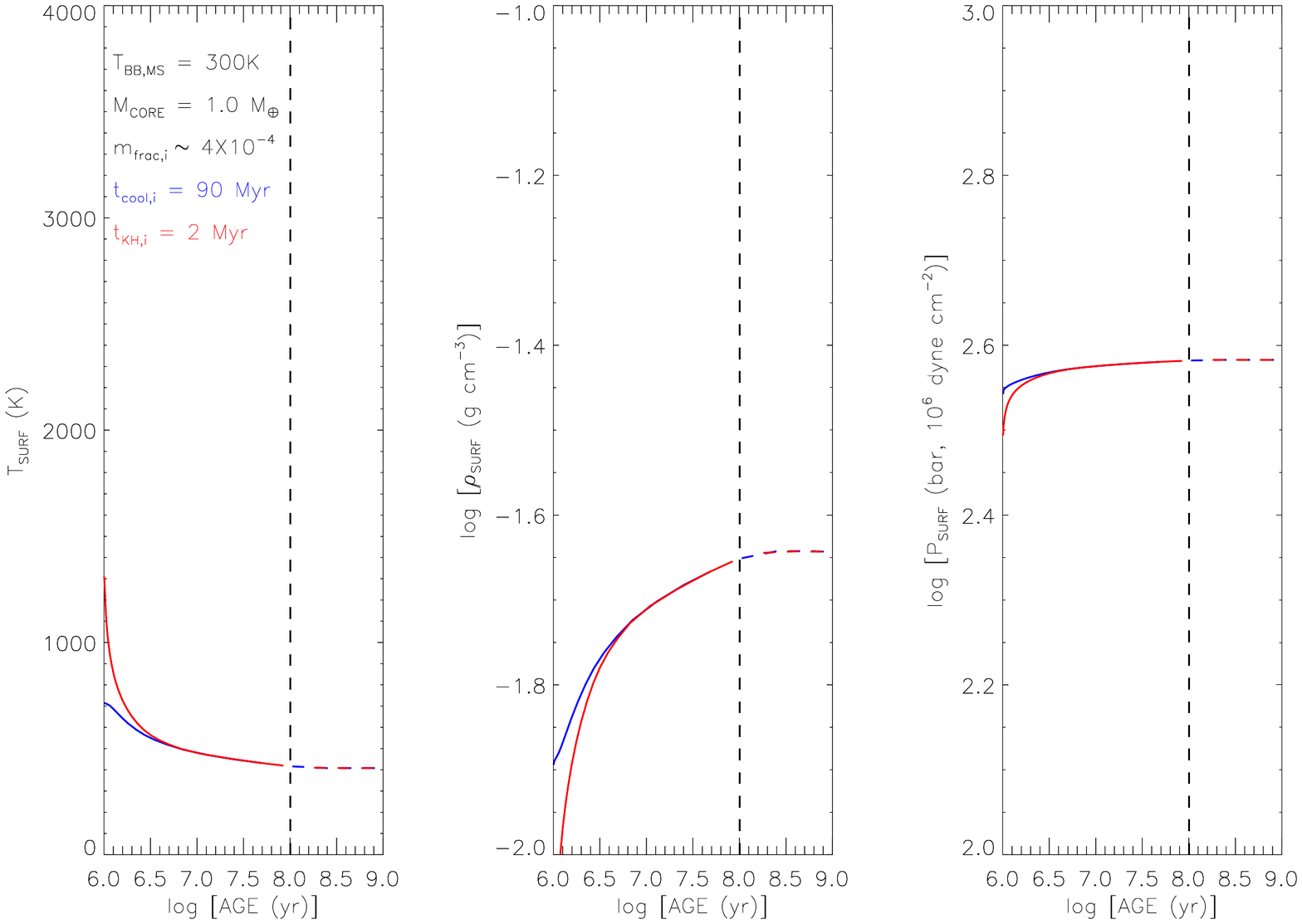}
\caption{Same `no-evaporation' calculation as in Figures~\ref{fig:0.8ne_300_4e-4}, but for $\mcore$\, = 1.0\,M$_{\oplus}$.}\label{fig:1.0ne_300_4e-4}
\end{figure}

\begin{figure}
\centering
\includegraphics[width=\columnwidth]{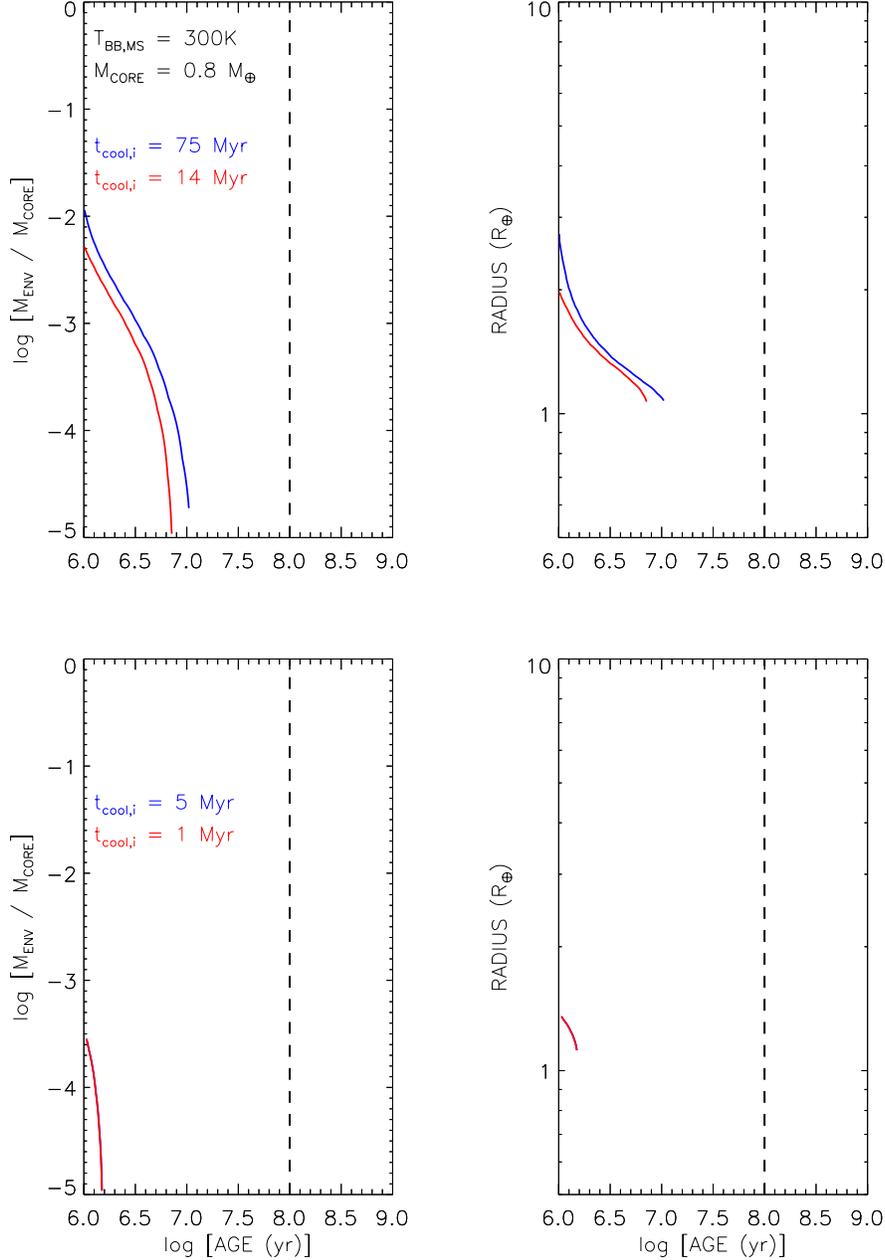}
\vspace{-1in}
\caption{Evolution of the envelope mass fraction (left panels) and planet radius (right) as a function of time for planets undergoing evaporation.  Once again, the {\it vertical dashed line} is at the median age of AD Leo, the {\it red curve} is for a ``hot-start'' model and the {\it blue curve} for a ``cold-start'' model. The planet is located at $\tbb1$\, = 300\,K (inner edge of HZ), with a core mass $\mcore$\, = 0.8\,M$_\oplus$. The top row is for planets with an initial H/He envelope mass-fraction of $m_{\rm frac,i}$ $\approx$ 10$^{-2}$, while the bottom row is for $m_{\rm frac,i}$ $\approx$ 4$\times$10$^{-4}$. The evolution of the ``hot-start'' and ``cold-start'' models is essentially indistiguishable in the bottom row. }\label{fig:0.8e_300}
\end{figure}

\begin{figure}
\centering
\includegraphics[width=\columnwidth]{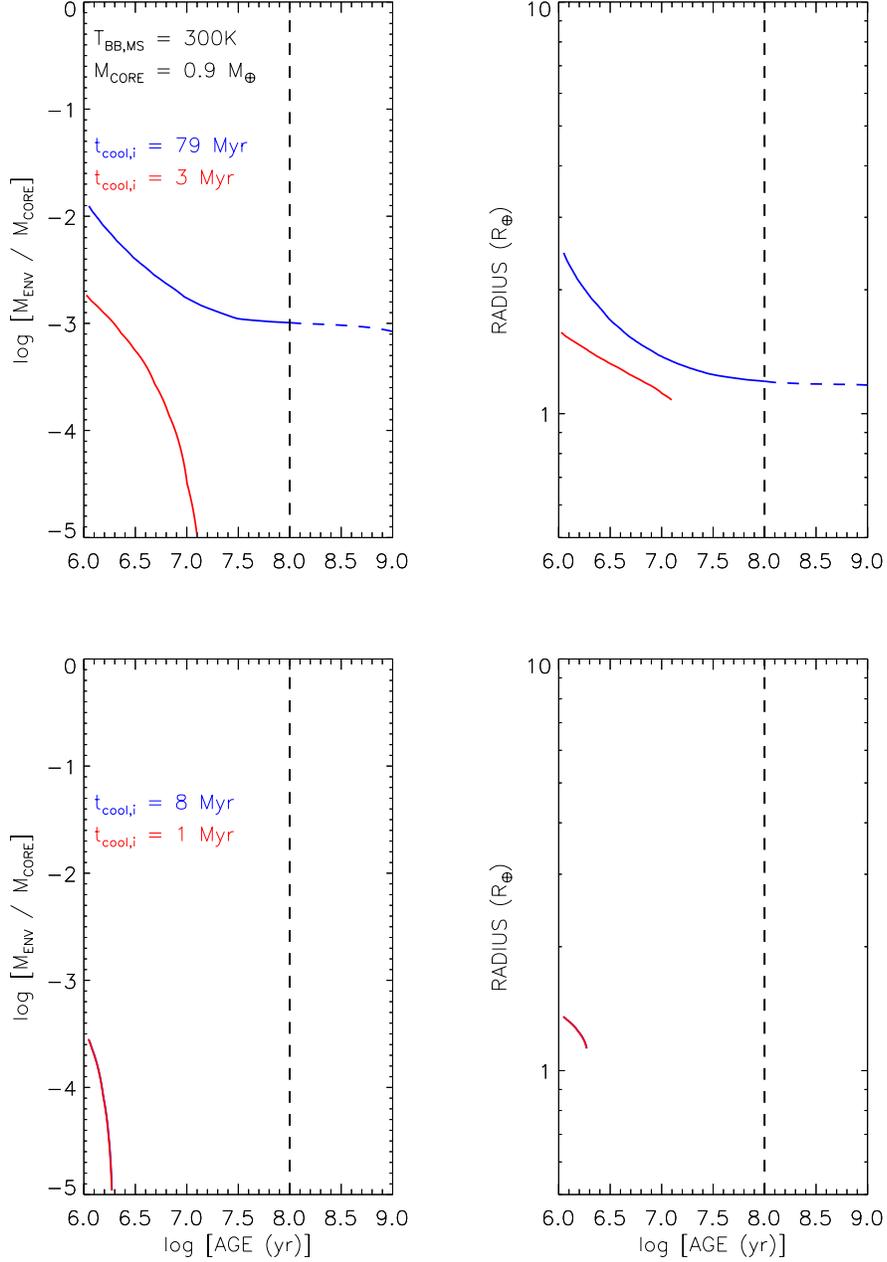}
\vspace{-1in}
\caption{Same as Figure~\ref{fig:0.8e_300}, but for $\mcore$\, =  0.9\,M$_\oplus$. Note that the ``hot-start'' model ({\it red curve}), for $m_{\rm frac,i}$ $\approx$ 10$^{-2}$ (top panels), undergoes ``boil-off'' \citep{ow15} at very early times, resulting in an almost instantaneous decrease in the initial envelope mass and planetary radius compared to the ``cold-start'' model (the initial boil-off phase is not plotted for clarity) - see text.}\label{fig:0.9e_300} 
\end{figure}

\begin{figure}
\centering
\includegraphics[width=\columnwidth]{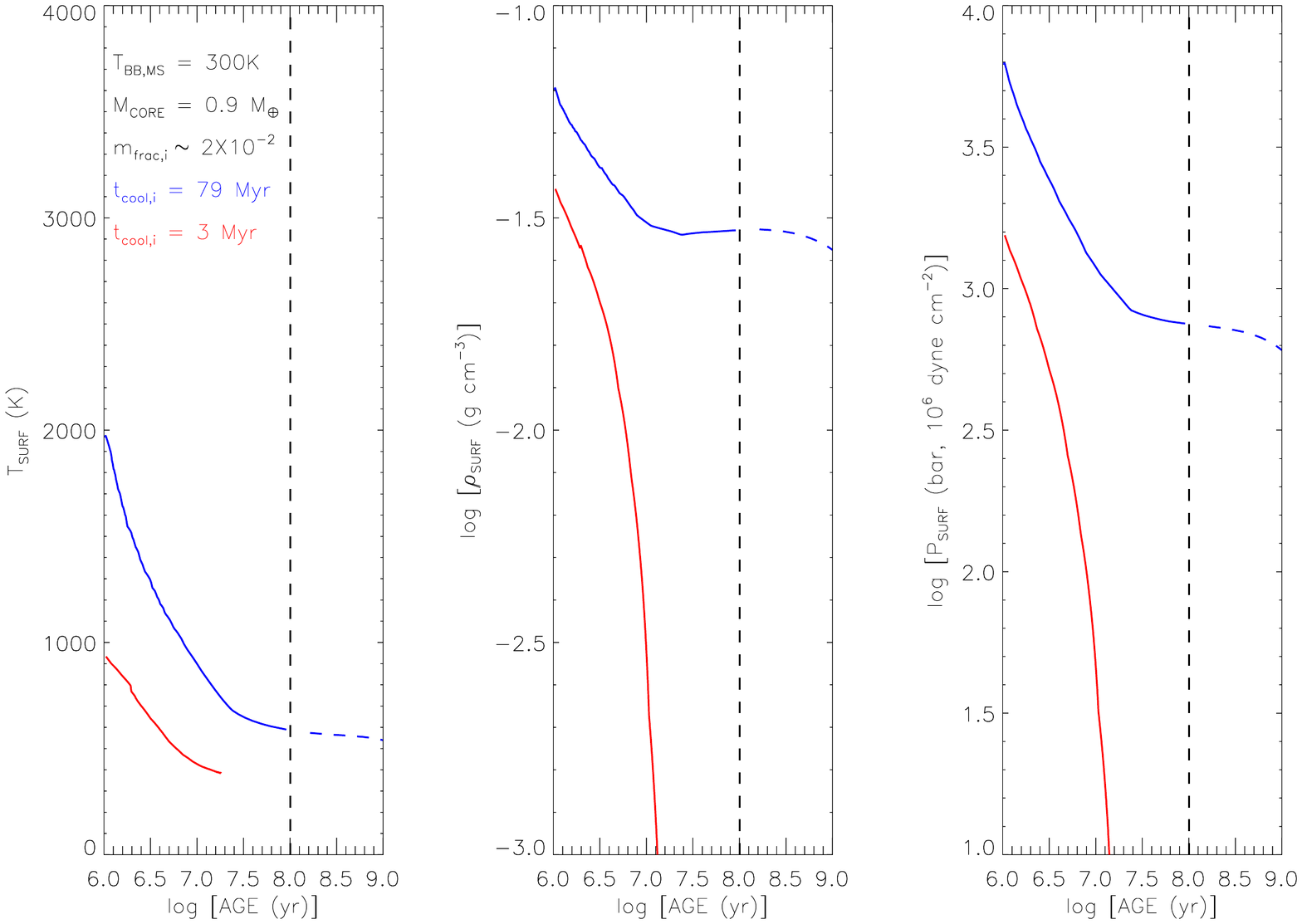}
\caption{Evolution of the surface temperature (left panel), density (middle) and pressure (right) as a function of time for evaporating planets corresponding to the top row of Figure~\ref{fig:0.9e_300} (i.e., planets at $\tbb1$\, = 300\,K, with $\mcore$\, =  0.9\,M$_\oplus$ and $m_{\rm frac,i}$ $\approx$ 10$^{-2}$).}\label{fig:0.9e_300_0.01}
\end{figure}

\begin{figure}
\centering
\includegraphics[width=\columnwidth]{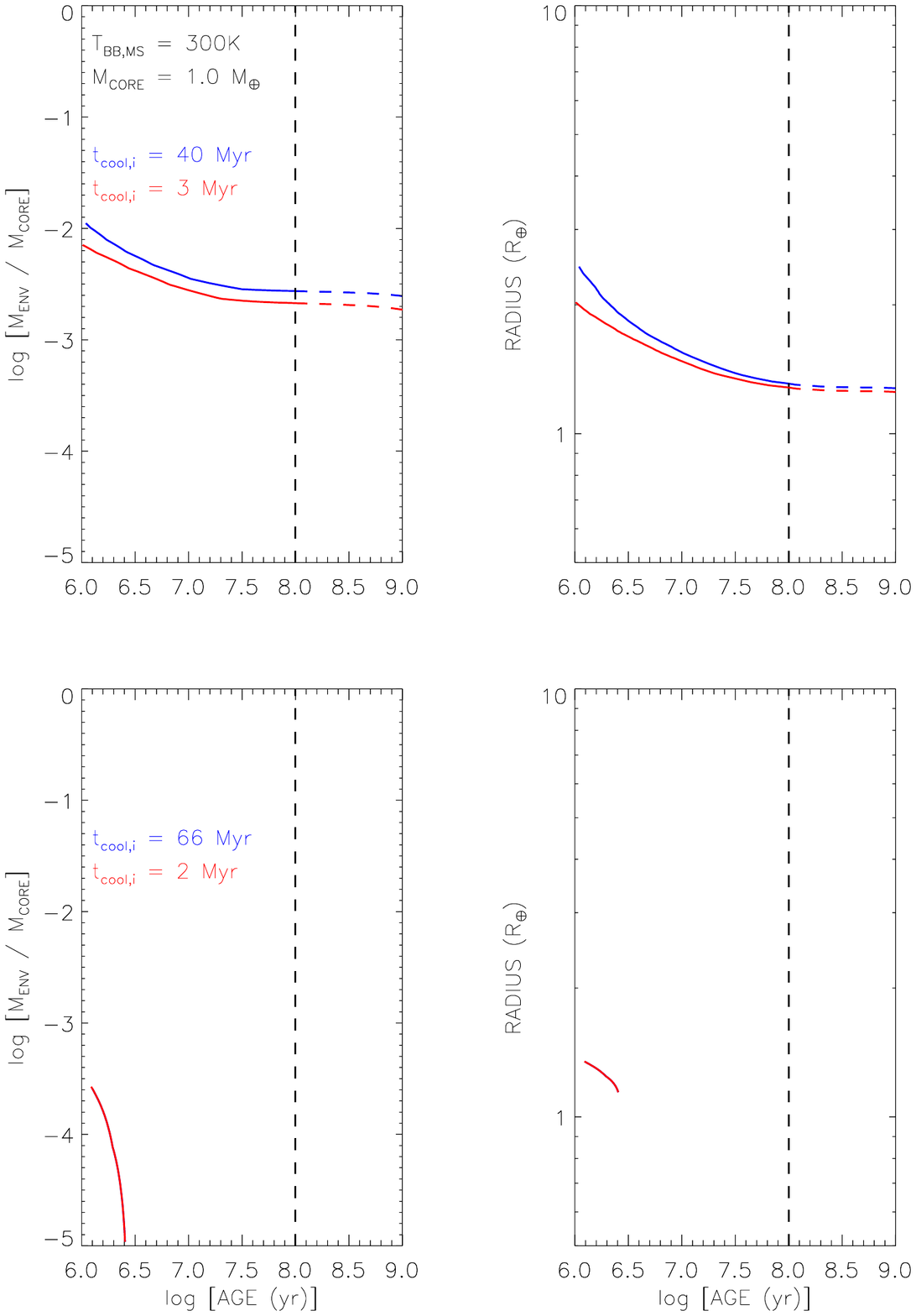}
\vspace{-1in}
\caption{Same as Figure~\ref{fig:0.8e_300}, but for $\mcore$\, =  1.0\,M$_\oplus$.}\label{fig:1.0e_300}
\end{figure}

\begin{figure}
\centering
\includegraphics[width=\columnwidth]{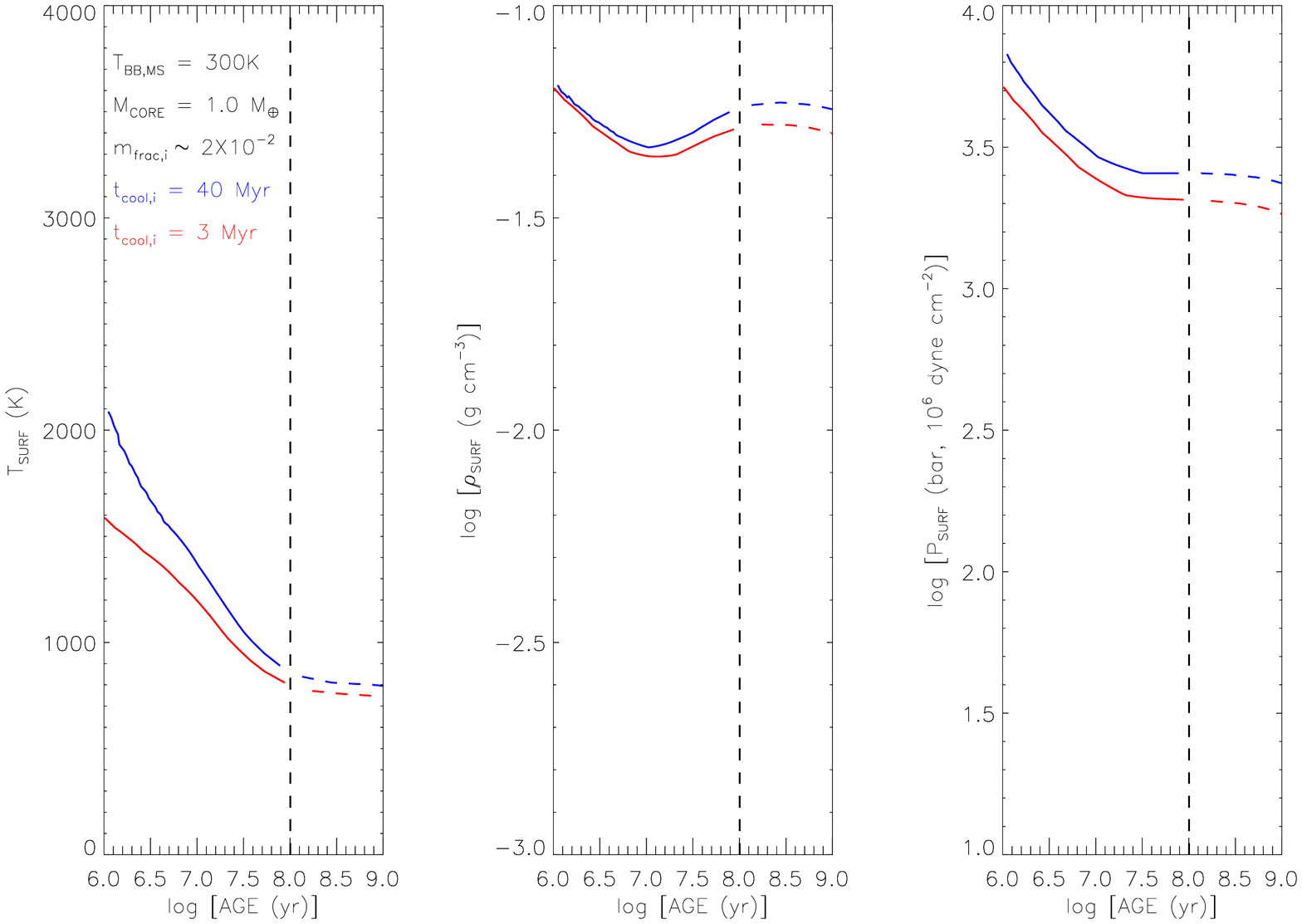}
\caption{Evolution of the surface temperature (left panel), density (middle) and pressure (right) as a function of time for evaporating planets corresponding to the top row of Figure~\ref{fig:1.0e_300} (i.e., planets at $\tbb1$\, = 300\,K, with $\mcore$\, =  1.0\,M$_\oplus$ and $m_{\rm frac,i}$ $\approx$ 10$^{-2}$).}.\label{fig:1.0e_300_0.01}
\end{figure}

\begin{figure}
\centering
\includegraphics[width=\columnwidth]{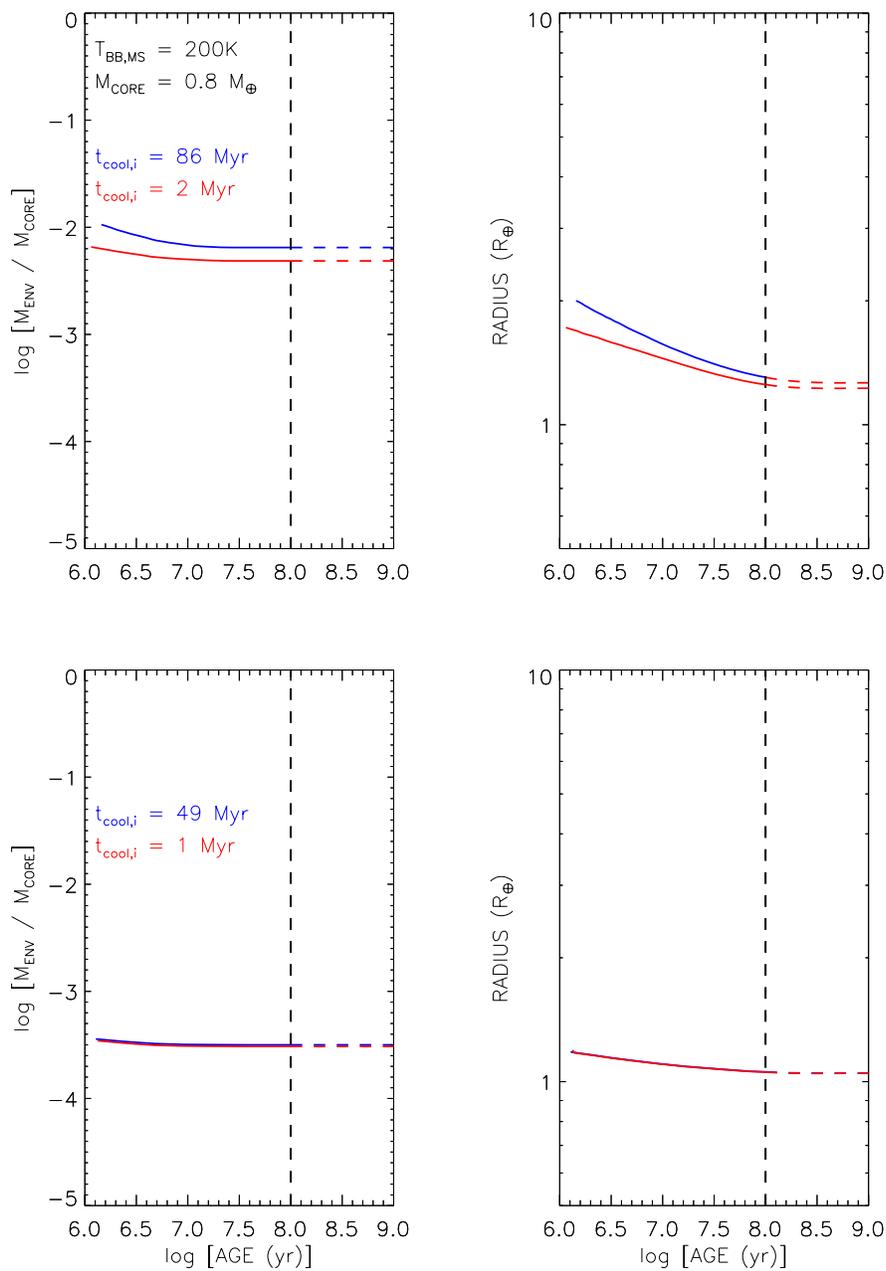}
\vspace{-1in}
\caption{Same as Figure~\ref{fig:0.8e_300}, with the same core mass $\mcore$\, = 0.8\,M$_{\oplus}$, but now for a radial location corresponding to $\tbb1$\, = 200\,K (i.e., outer edge of HZ).}\label{fig:0.8e_200}
\end{figure}

\begin{figure}
\centering
\includegraphics[width=\columnwidth]{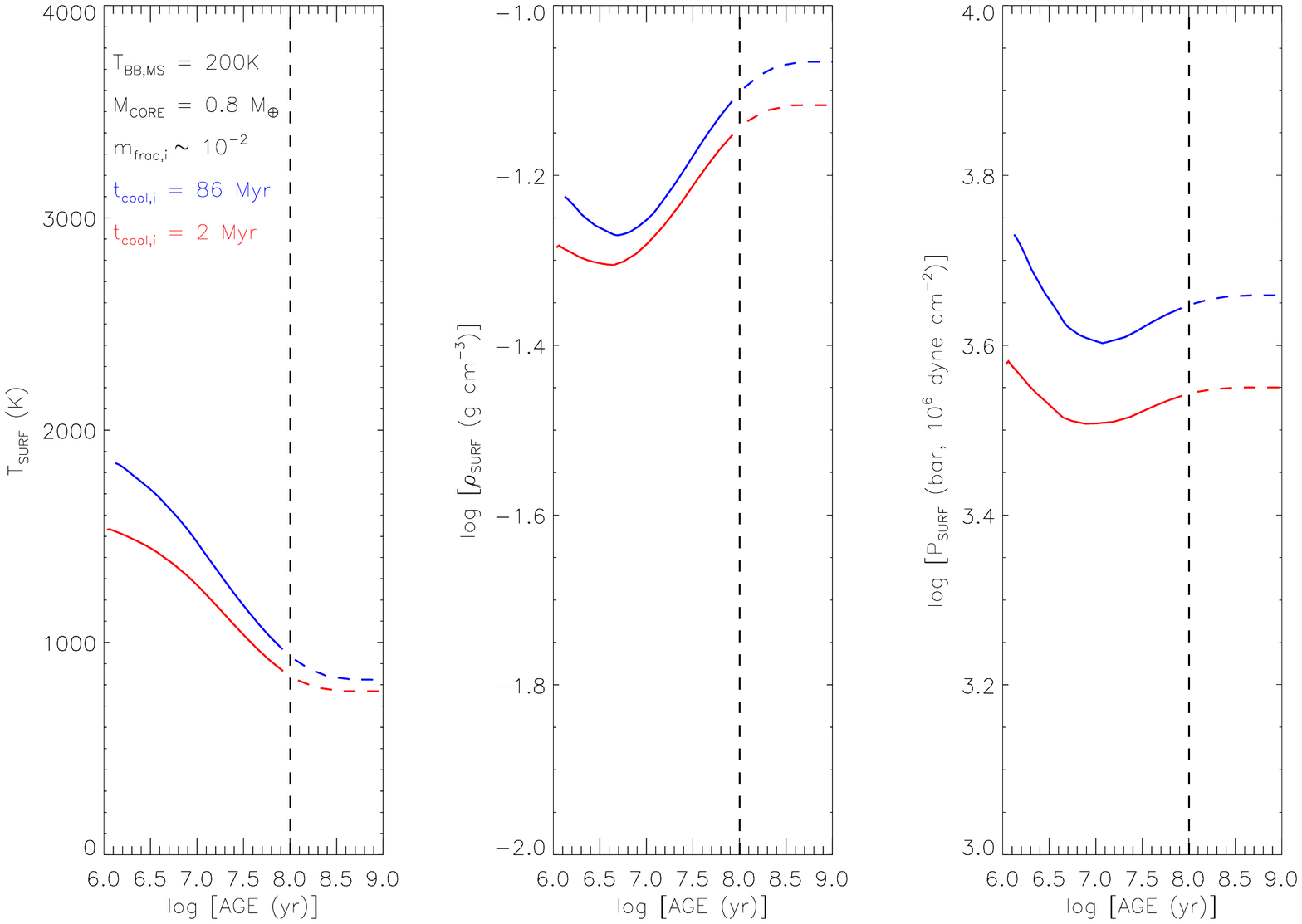}
\caption{Evolution of the surface temperature (left panel), density (middle) and pressure (right) as a function of time for evaporating planets corresponding to the top row of Figure~\ref{fig:0.8e_200} (i.e., planets at $\tbb1$\, = 200\,K, with $\mcore$\, =  0.8\,M$_\oplus$ and $m_{\rm frac,i}$ $\approx$ 10$^{-2}$).}\label{fig:0.8e_200_0.01}
\end{figure}

\begin{figure}
\centering
\includegraphics[width=\columnwidth]{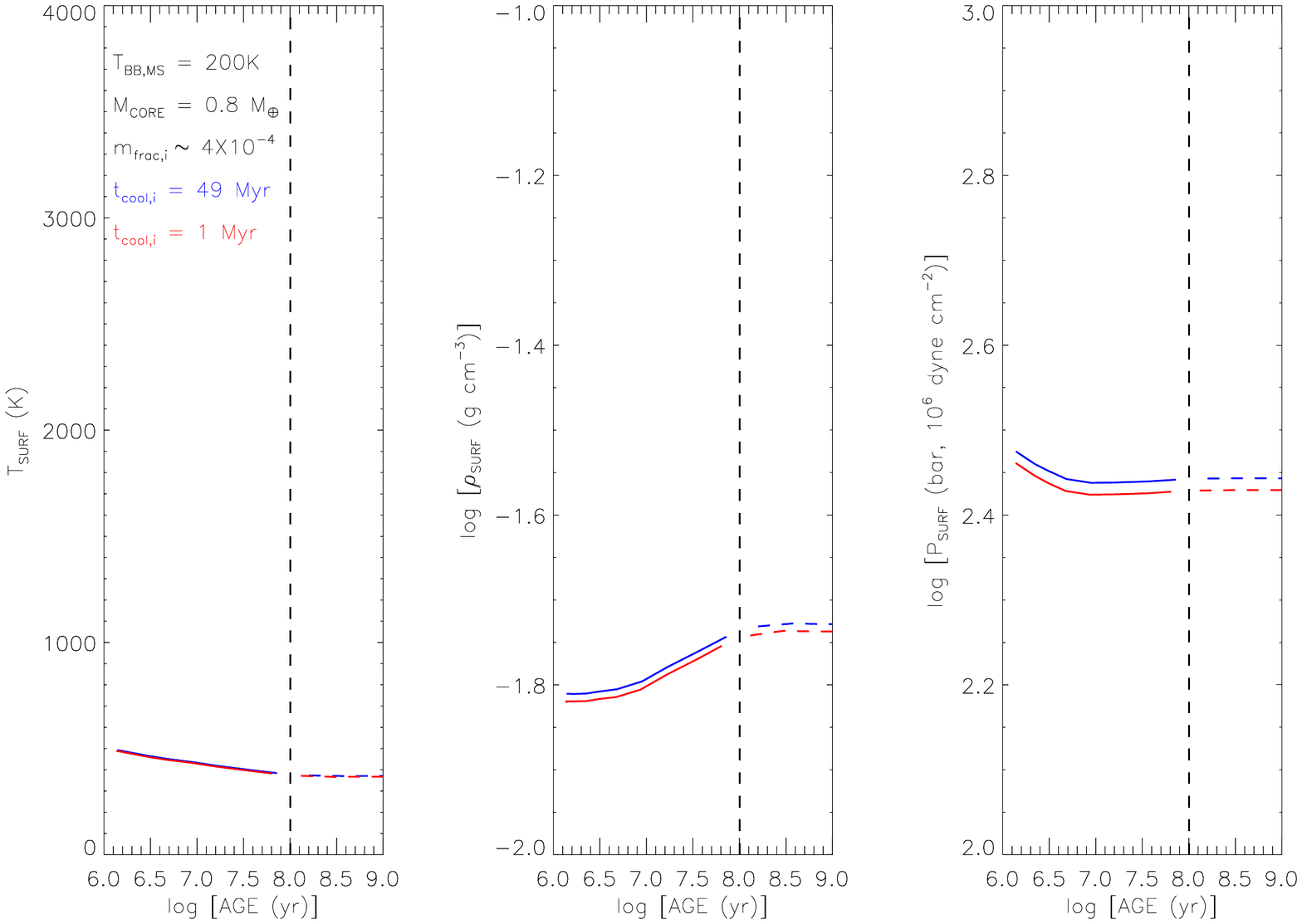}
\caption{Evolution of the surface temperature (left panel), density (middle) and pressure (right) as a function of time for evaporating planets corresponding to the bottom row of Figure~\ref{fig:0.8e_200} (i.e., planets at $\tbb1$\, = 200\,K, with $\mcore$\, =  0.8\,M$_\oplus$ and $m_{\rm frac,i}$ $\approx$ 4$\times$10$^{-4}$).}\label{fig:0.8e_200_4e-4}
\end{figure}

\begin{figure}
\centering
\includegraphics[width=\columnwidth]{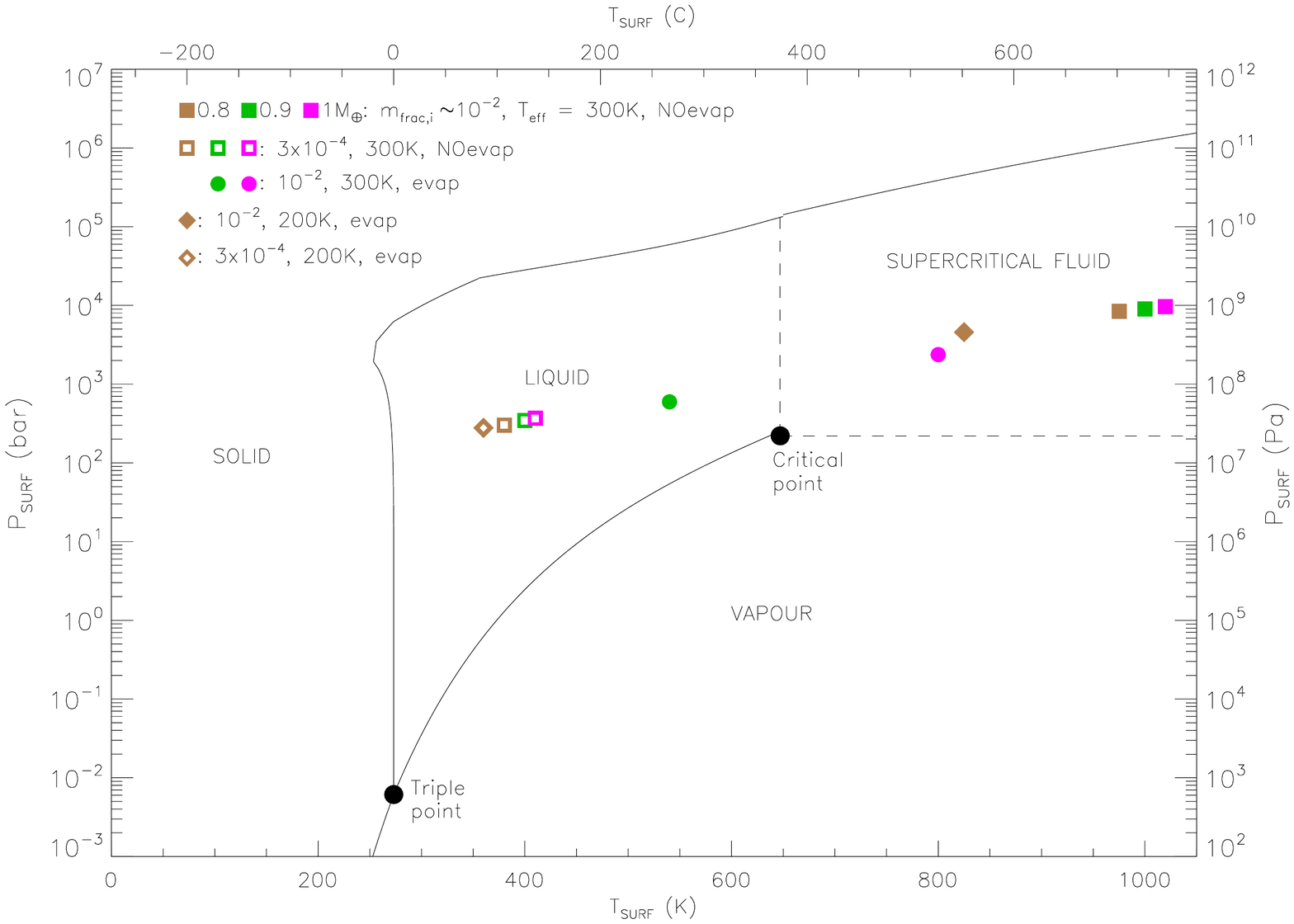}
\caption{Pressure-Temperature phase diagram of water. The points show the 1~Gyr surface conditions of those of our planets that have retained some of their initial H/He envelopes.}\label{fig:phasediagram}
\end{figure}

\section{Discussion}
The central result of the preceding analysis is that solid cores made of 2/3 rock + 1/3 iron, with mass $\gtrsim$1\,\me, cannot lose enough of their envelopes to become habitable in the classical HZ of M dwarfs, if they are born with significant H/He envelopes. The fundamental reason is that hydrodynamic escape is quenched in such planets while a sizeable portion of the H/He envelope still lingers, and subsequent Jeans escape -- which can at best extract a fractional mass of $\sim$10$^{-5}$ over a Gyr -- is too feeble to remove this remainder. Only cores $<$1\,\me\, (with the precise limiting mass depending on where in the HZ one is) can undergo hydrodynamic escape all the way down to their surface in spite of having initial H/He mass-fractions of $\sim$1\%, and may thus be habitable (either due to a tenuous remnant H/He envelope, or a similarly wispy secondary atmosphere) in the classical HZ of M dwarfs. 

\subsection{Implications for Habitable Planets around M Dwarfs}
To summarise: {\it Kepler} data imply that small (terrestrial-size) planets are ubiquitous at small orbital separations, with nearly every star (statistically speaking) hosting one such ``Kepler'' planet; around M dwarfs, a sizeable fraction of these close-in planets reside within the classical HZ (see also further below). Concurrently, both observations and theory indicate that such planets are usually born with H/He envelopes with a mass-fraction $\gtrsim$1\%, and our calculations imply that planets with cores comprising 2/3 rock and 1/3 iron, with core masses $\gtrsim$1\,\me\, and initial H/He envelope mass-fractions $\gtrsim$1\%, cannot lose enough of their envelopes to be habitable within the classical HZ of M dwarfs.  With these considerations, we propose three possible classes of solid-core habitable planets around M dwarfs:

\noindent {\it (1) Sub-Earth mass rock/iron-core planets within the classical HZ}: We found above that sub-Earth mass rock/iron cores -- with mass $\lesssim$0.9\,\me\, near the inner edge of the classical HZ of M dwarfs, and $<$0.8\,\me\, near the outer edge -- can be stripped of $\sim$1\% initial H/He envelopes within a Gyr. Such $\lesssim$Venus-mass planets may therefore be habitable within the classical HZ of M dwarfs, if they acquire secondary atmospheres like the solar system terrestrial planets (the HZ results of \citet{kopparapuetal2013} can then be applied to such planets). 

\noindent {\it (2) Sub-Earth to super-Earth mass ice-core planets within the classical HZ}: Our discussion so far has focussed on rock/iron cores. Naively, the results from {\it (1)} above should apply to ice cores as well (but see below), except for a higher threshold mass (the limiting planet mass below which $\sim$1\% initial H/He envelopes will be stripped). This is easily seen by examining Fig.\,4, where the threshold mass is defined as the intersection of the Kn$_s$=1 line and the locus of a core with fixed density. For a given core mass, an ice core, with a density $\sim$10 times smaller than a rock/iron one, will have the latter locus shifted to the right (i.e., to a larger radius) by a factor of 10$^{1/3}\sim$2, resulting in a threshold mass larger by a factor of $\sim$2 as well from Fig.\,4. Thus, to zeroeth order, one expects ice-core planets, with a somewhat (factor of $\sim$2) larger limiting mass than rock/iron core ones, to also be habitable within the classical HZ of M dwarfs. However, a correct analysis of evaporation here must also account for the opacity of steam (liberated from the core's surface) in addition to H/He, which may strongly affect the outcome (since oxygen -- in this case within H$_2$O molecules -- is a strong X-ray absorber). This will be the subject of future work. Note also that icy cores can only form beyond the ice-line, situated at $\sim$1.5\,AU for a $\sim$0.4\,\msun\, M dwarf at an age of order a Myr\footnote{Using the formula supplied by \citet{ida08} for the ice-line orbital separation: $a_{ice} = 2.7 (L_{\ast}/L_{\odot})^{1/2}$\,AU; with a stellar bolometric luminosity of $L_{\ast} \sim 0.3$\,$L_{\odot}$ for a 0.4\,\msun\, star at an age of 1.5\,Myr, from the evolutionary tracks of \citet{baraffe98}}. Thus ice-cores must migrate inwards for the above discussion to be pertinent. 

\noindent {\it (3) Planets formed after gas-disk dispersal}: Finally, our entire analysis in this paper is for planets that form while a significant amount of gas still remains in the surrounding primordial disk, allowing them to accrete relatively massive H/He envelopes; {\it Kepler} data, as discussed, imply that such planets are ubiquitous. However, the terrestrial planets in our own solar system evince signatures of having coalesced after the gas disk dispersed (as the radiometricly determined age of the earth -- \citealt{manhes80} -- is estimated to be $50-100$~Myr younger than that of the solar-system, e.g. \citealt{bouvier10}, well after the gas disc would have dispersed). If solid-core planets around M dwarfs can form in a similar fashion, then they might well be habitable in the classical HZ of M dwarfs (and slightly interior / exterior to it depending on planetary mass); the work by \citet{kopparapuetal2013}, examining solely secondary atmospheres, is more directly applicable in that case than this paper.    

We note that there is, prima facie, another interesting possibility.  While we find that rock/iron cores $\gtrsim$1\,\me\, with $\sim$1\% initial H/He envelopes retain too large a fraction of these envelopes to be habitable within the classical HZ of M dwarfs, they {\it will} be stripped of such atmospheres if their orbits are much smaller. Simultaneously, \citet{kopparapuetal2013} show that the inner boundary of the HZ (set by either the moist greenhouse or runaway greehouse effect) for such {\it super}-Earth mass planets will be closer to the star than for Earth-mass ones (where the latter are used, by definition, to calculate the classical HZ limits). Therefore, one might expect an overlap in space, interior to the classical HZ, between where a super-Earth of given mass can be stripped of a $\sim$1\% H/He envelope, and where it will be habitable with a secondary atmosphere before greenhouse effects become overwhelming; super-Earths would then be habitable within this overlap region. Unfortunately, such an overlap is implausible. For solar-type stars, \citet{kopparapuetal2013} show that the inner boundary of the HZ moves inwards from 0.99\,AU for a 1\,\me\, planet to 0.94\,AU for a 10\,\me\, one; i.e., only a very small shift of 0.05\,AU for a factor of 10 increase in planetary mass. The effect should be similar around M dwarfs. This tiny decrease in orbital separation will have negligible impact on the incident X-ray flux and thus on the mass-loss rates; as such, we do not expect any HZ for super-Earths interior to the classical HZ around M dwarfs. Conversely, for sub-Earth mass planets, the inner boundary of the HZ moves slightly {\it outwards}, as \citet{kopparapuetal2013} show; thus, we do not expect any HZ for these planets interior to the classical HZ either.

\begin{figure}
\centering
\includegraphics[width=\columnwidth]{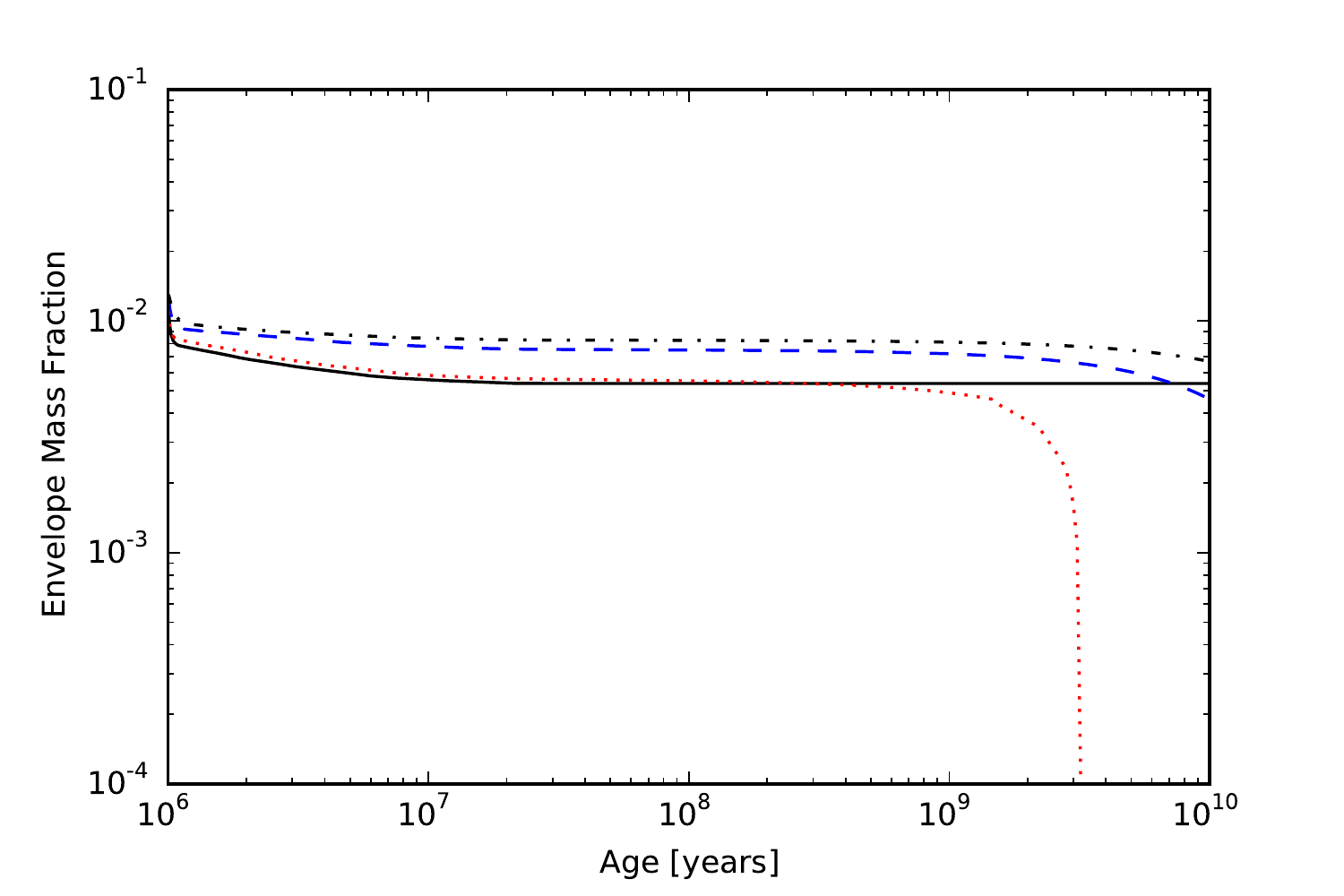}
\caption{Envelope mass fraction evolution for our 1~M$_\oplus$ planet at the inner edge of the HZ. The {\it solid} curve shows the evolution with the mass-loss rates calculated in this work, while the {\it dot-dashed}, {\it dashed} and {\it dotted} curves show mass-loss rates calculated in the ``energy-limited'' framework with efficiencies of $\eta$ = 5, 10 and 25\% respectively. Note that we have assumed that the X-ray flux remains saturated over the entire evolution: even in this extreme case, the flow still becomes non-hydrodynamic (ballistic) at late times in our model, preventing runanway mass loss; this is not true of the energy-limited cases.}\label{fig:EL_compare}
\end{figure}

There are two main ingredients that determine the actual frequency of habitable planets around M dwarfs. The first, as we have shown, is a rigorous treatment of evaporation, using hydrodynamic models that account for radiative cooling as well as the transition to Jeans escape, and including the thermal evolution of the planet. A simplistic ``energy-limited'' formalism, that moreover assumes that the escape is always hydrodynamic (e.g., \citealt{luger15a} for M dwarfs), leads to a gross overestimation of evaporative mass-loss rates, and thereby an overly optimistic appraisal of the fraction of planets that can be rendered habitable by stripping their primordial H/He envelopes. {\rc Our results are in direct conflict with those of \citet{luger15a}. The latter authors' calculations suggest that in the majority of cases, M dwarfs can completely strip envelopes with mass fraction $\lesssim$\,10\% (and even larger ones in some cases) at the inner edge of the HZ, and mass fractions $\lesssim$\,1\% at the outer edge of the HZ, from 1--2 M$_\oplus$ cores. In other words, they conclude that evaporation can generate a plethora of potentially habitable Earth-mass planets around M-dwarfs. Our results indicate that this is not the case: it is the simplifications made by \citet{luger15a} thet lead them to conclude otherwise. The main reason is their choice of a constant efficiency energy-limited mass-loss prescription. As we have discussed throughout this work, at late times when the XUV flux declines and the envelope mass fraction drops below $\sim$\,1\% (i.e., when the planet's radius shrinks rapidly with decreasing envelope mass), the efficiency of hydrodynamic mass-loss decreases significantly \citep[c.f.][]{ow13}; furthermore, the flow transitions to non-hydrodynamic (Jeans escape) at late times. Neglecting these effects, \citet{luger15a} find that their planets can undergo runaway mass-loss at ages $> 0.1-1$~Gyr; including these effects, however, we find that a $\sim$1\,M$_\oplus$ planet can only ever undergo runaway mass-loss at early times ($<0.1$~Gyr), when the flux is high enough to permit hydrodynamic evaporation. In fact, if we adopt an energy-limited mass-loss rate with no hydrodynamic cut-off at low fluxes, we indeed recover the runaway mass-loss at late times found by \citet{luger15a}. This is demonstrated in Fig.~\ref{fig:EL_compare}, where we compare the results of our calculations (solid line) to those for the energy-limited case with efficiencies ($\eta$) of 5\% (dot-dashed), 10\% (dashed) \& 25\% (dotted); note that \citet{luger15a}'s default choice is $\eta$ = 30\%. The plot shows that, while our planet loses negligible mass at late times due to the hydrodynamic cut-off, the energy-limited cases can undergo unphysical runaway mass-loss at these ages. We note that while an efficiency of $10\%$ produces a good match to the final envelope mass-fraction at 10~Gyr in this {\it particular} example, this is {\it not} a generally applicable result: indeed, the plot shows that the 10\% case (dashed line) is also starting to enter a run-away phase at the end of the calculation. In general, there is no single efficiency value that can mimic the trends in evaporation in our results, especially the hydrodynamic cut-off, which turns out to be extremely important for evaporation and the question of potential habitability. Using an energy-limited formalism to determine evaporation rates when the flow is close to transitioning from hydrodynamic to non-hydrodynamic can lead to rather poor predictions.} Finally, we note that our prediction that planets $\gtrsim$1\,\me\, will retain most of their initial voluminous H/He envelopes, in the classical HZ of M dwarfs, should be directly testable with TESS, which will target a large number of M dwarfs. 

The second ingredient that controls the actual prevalence of habitable planets around M dwarfs is the frequency of their planets as a function of planetary mass, at the small end of the planetary spectrum. A recent careful analysis by \citet{morton14} indicates that the occurrence of M dwarf planets continues to increase to planetary radii below 1\,\re, without any strong evidence for a downturn in the planet radius function that was suggested by previous studies. If this is true, then it bodes well for the frequency of potentially habitable planets orbiting M dwarfs, since the smallest (lowest-mass) planets will be most easily stripped of their initial H/He envelopes. Indeed, applying their improved calculations to previous work by \cite{dressing13} and \cite{kopparapu2013}, \citet{morton14} already show that the frequency of Earth-size planets in the HZ of M dwarfs -- calculated by \cite{dressing13} to be $\sim$0.15 planets per star -- increases to $\sim$0.25--0.8, a remarkable fraction. If this frequency continues to increase towards smaller planets, as \cite{morton14} suggest, then evaporation can indeed create a bonanza of potentially habitable planets around these cool red dwarfs. However, the latter results are based on a quite limited sample; they need to be verified and sharpened with more data from the K2 mission, as well as upcoming ones such as TESS.

\section{Summary}

\noindent Our key results are as follow:

\noindent 1) Jeans escape can only remove a very small H/He mass-fraction from low-mass exoplanets in the HZ of M dwarfs ($\sim$10$^{-5}$ over a Gyr): hydrodynamic evaporation is essential for removing significant amounts of H/He. 

\noindent 2) Previously derived mass-loss rates, based on the ``energy-limited'' (or ``locally energy-limited'') formalism, and assuming moreover that escape is always in the hydrodynamic limit, are significant overestimations. Our improved model, accounting for both radiative losses and the transition from hydrodynamic to Jeans escape, and including the thermal evolution of the planet, yields much lower mass loss rates.

\noindent 3) In particular, for cores made of 2/3 rock + 1/3 iron, we find that only sub-Earth mass cores -- $\lesssim$0.9\,\me\, near the inner edge of the classical HZ of M dwarfs, and $<$0.8\,\me\, near the outer edge -- can be stripped of $\sim$1\% initial H/He envelopes. Rock/iron cores with mass $\gtrsim$1\,\me, and born with H/He envelope mass-fractions $\gtrsim$1\%, cannot lose sufficient envelope mass to become habitable within the classical HZ of M dwarfs. Our prediction that $\gtrsim$1\,\me\, cores in the HZ of M dwarfs should still be shrouded by their voluminous natal H/He envelopes over Gyr timescales will be directly testable with TESS.       

\noindent 4) We propose three classes of potentially habitable planets in the HZ of M dwarfs: {\it (i)} planets with sub-Earth mass rock/iron cores: since they can be stripped of 1\% H/He envelopes within a Gyr, they may be habitable if they can acquire suitable secondary atmospheres, like solar system terrestrial planets; {\it (ii)} planets with ice-cores, with an upper limit to the core mass somewhat higher than the sub-Earth mass for rock/iron cores, may also be able to lose their primordial H/He envelopes and thus become habitable; however, steam opacity must be included in the evaporation calculations to verify this possibility; and {\it (iii)} planets (possibly like the terrestrial ones in the solar system) that form after gas disk dispersal, and thus have only tenuous secondary atmospheres, with very little primordial H/He.

\acknowledgements
We thank the referee for an instructive report. We are grateful to Barbara Ercolano, Ren\'e Heller and Renyu Hu form comments on the manuscript. We are indebted to Jorge Sanz-Forcada for supplying the synthetic XUV spectrum of AD Leo, which enabled a more realistic analysis of evaporation around M dwarfs than possible with the scaled-solar spectra used in previous work, and for comments on the manuscript. We are also very grateful to Martin Chaplin for providing the equations for the phase diagram of water. JEO acknowledges support provided by NASA through Hubble Fellowship grant HST-HF2-51346.001-A awarded by the Space Telescope Science Institute, which is operated by the Association of Universities for Research in Astronomy, Inc., for NASA, under contract NAS 5-26555. SM acknowledges the support of STFC-UK Consolidated Grant ST/K001051/1.



\end{document}